\documentclass[12pt,onecolumn]{IEEEtran}
\usepackage[dvips,final]{graphicx}
\usepackage{latexsym}
\usepackage{amssymb}
\usepackage[cmex10]{amsmath}
\usepackage{amsthm}
\usepackage{epstopdf}
\usepackage{color}
\usepackage{bm}
\usepackage{bbm}
\usepackage{array}
\usepackage{balance}
\usepackage[caption=false,font=footnotesize]{subfig}
\usepackage{cite}
\usepackage{algorithmic}
\usepackage{algorithm}

\newtheorem{proposition}{Proposition}
\newcommand{\argmax}{\operatornamewithlimits{arg\,max}}

\title{Reconstruction of Network Coded Sources From Incomplete Datasets}

\author{Eirina~Bourtsoulatze,~\IEEEmembership{Member,~IEEE,}
	     Nikolaos~Thomos,~\IEEEmembership{Member,~IEEE,} 
              and~Pascal~Frossard,~\IEEEmembership{Senior Member,~IEEE} %
\thanks{E. Bourtsoulatze is with the Communication and Distributed Systems (CDS) laboratory, Institute of Computer Science and Applied Mathematics, University of Bern, Bern CH-3012, Switzerland (e-mail: bourtsoulatze@iam.unibe.ch).
N. Thomos is with the Department of Computer Science and Electronic Engineering, University of Essex, Colchester CO4 3SQ, U.K. (e-mail: nthomos@essex.ac.uk).
P. Frossard is with the Signal Processing Laboratory 4 (LTS4), Ecole Polytechnique F\'{e}d\'{e}rale  de Lausanne (EPFL), Lausanne CH-1015, Switzerland (e-mail: pascal.frossard@epfl.ch).}%
\thanks{This work has been supported by the Swiss National Science Foundation under grants 200021-138083 and PZ00P2-137275.}}

\begin{document}

\maketitle

\begin{abstract}

In this paper, we investigate the problem of recovering source information from an incomplete set of network coded data. We first study the theoretical performance of such systems under {\em maximum a posteriori} (MAP) decoding and derive the upper bound on the probability of decoding error as a function of the system parameters. We also establish the sufficient conditions on the number of network coded symbols required to achieve decoding error probability below a certain level. We then propose a low complexity iterative decoding algorithm based on message passing for decoding the network coded data of a particular class of statistically dependent sources that present pairwise linear correlation. The algorithm operates on a graph that captures the network coding constraints, while the knowledge about the source correlation is directly incorporated in the messages exchanged over the graph. We test the proposed method on both synthetic data and correlated image sequences and demonstrate that the prior knowledge about the source correlation can be effectively exploited at the decoder in order to provide a good reconstruction of the transmitted data in cases where the network coded data available at the decoder is not sufficient for exact decoding.

\end{abstract}

\begin{keywords}
Network coding, correlated sources, message passing, approximate decoding.
\end{keywords}


\section{Introduction}
\label{sec:intro}

The emergence of new network architectures and the growth of resource demanding applications have created the need for novel data delivery mechanisms that are able to efficiently exploit the network diversity. This has favored the emergence of the network coding research field \cite{Ahlswede00} that introduces the concept of in-network data processing with the aim of improving the network performance. Extensive theoretical studies have revealed the great potential of network coding as being a building block for efficient data delivery. It has been shown for example that one can achieve the maximum network flow in the multicast communication scenario by randomly combining the incoming data in the network nodes \cite{RandomizedNC03}. 

While many research efforts have focused on the design of network codes, only few works have addressed the problem of recovering the network coded data when the available network coded information is not sufficient for perfect decoding. The assumptions made by the majority of network coding algorithms is that the source data is decoded upon collecting a sufficient number of network coded packets, so that the original data can be perfectly reconstructed using exact decoding methods such as Gaussian elimination. This imposes significant limitations on the application of network coding in real systems; due to dynamics in the network, such as bandwidth variations or data losses, there is no guarantee that the required amount of network coded data will reach the end user in time for decoding. Under these conditions, state of the art methods can only provide an all-or-nothing performance, with the source data being either perfectly recovered or not recovered at all. Besides its negative impact on the delivered data quality, this also influences the network performance since network resources may actually be used to deliver useless network coded data. In such cases, approximate reconstruction of the source data from incomplete network coded data is surely worthful, even if perfect reconstruction cannot be achieved.

In this paper, we build on our previous work \cite{BourtsouNetCod12} and we investigate the problem of decoding the original source data from an incomplete set of network coded data with help of source priors. We consider correlated sources that generate discrete symbols, which are subsequently represented by values in a finite algebraic field and transmitted towards the receivers for further processing. The intermediate network nodes perform randomized linear network coding. Then, given a set of network coded symbols, the decoder provides an approximate reconstruction of the original source data. We first study the theoretical performance of an optimal {\em maximum a posteriori} decoder. We provide an upper bound on the probability of erroneous decoding as a function of the number of network coded symbols available at the decoder, the finite field size that is used for data representation and network coding, and the joint probability mass function of the source data. We also establish the lower bound on the required number of network coded symbols for achieving decoding error probability below a certain value. These bounds provide useful insights into the performance limits of the considered framework. 

Next, we propose a constructive solution to the problem of approximately decoding a class of statistically dependent sources that present linear pairwise correlation. We design an iterative decoding algorithm based on message passing, which jointly considers the network coding constraints given in the finite field and the correlation constraints that are expressed in the real field. We test our decoding algorithm on two sets of correlated sources. In the first case, synthetic source symbols are generated by sampling a multivariate normal distribution. In the second case, the source data corresponds to signals obtained from correlated images in a video sequence. The results demonstrate that the original source data can be efficiently decoded in practice from an incomplete set of network coded symbols by using a simple correlation model expressed as a correlation noise between pairs of sources. 

Note that, though the framework that we examine bears some resemblance to the distributed source coding problem and can be viewed as an instance of in-network compression, our primal objective is not to design a data compression algorithm. We rather aim at providing an approximate reconstruction of the source data in cases where the source and network coding have not been jointly optimized, and where the complete set of network coded data required for exact decoding by Gaussian elimination is not available at the decoder. Furthermore, it should be noted that distributed source coding schemes require that either the knowledge of the correlation between sources is available at each source in order to determine the optimal encoding rates or, that a feedback channel exist for adapting the source rates according to the progress of the decoder. This framework is very different from the settings considered in this paper, where network coding is independent of the source statistics, and where the decoder tries to recover information without direct communication with the sources. 

The rest of the paper is organized as follows. In Section \ref{sec:relatedwork}, we present an overview of the related work. We describe the network coding framework in Section \ref{sec:framework}. In Section \ref{sec:error}, we analyze the performance of a {\em maximum a posteriori} (MAP) decoder by giving an upper bound on the decoding error probability and the sufficient conditions on the number of network coded symbols required to achieve decoding error probability below a certain value. Next, in Section \ref{sec:decoding}, we propose a practical iterative decoding algorithm based on message passing for decoding network coded correlated data from an incomplete set of network coded symbols. In Section \ref{sec:results}, we illustrate the performance of the proposed decoding algorithm through simulations with synthetic data and with sets of correlated images obtained from video sequences. Section \ref{sec:conclusions} concludes this paper.


\section{Related work}
\label{sec:relatedwork}

The problem of decoding from an incomplete set of network coded data refers to the problem of reconstructing the source data from a set of network coded symbols that is not large enough for uniquely determining the values of the source symbols. When linear network coding in finite field is considered, this translates into the problem of recovering the source symbols from a rank deficient system of linear equations in a finite field. This problem resembles classical inverse problems \cite{TarantolaInvProb} in signal processing, which aim at recovering the signal from an incomplete set of observed linear measurements. However, while the latter are successfully solved using regularization techniques  \cite{regularizationreview}, the same methods cannot be applied to the source recovery problem in finite domains, due to the different properties of the finite field arithmetic, {\em e.g.} the cyclic property. 

Source recovery from an insufficient number of network coded data is an ill-posed problem and requires additional assumptions on the structure of the solution, {\em e.g.} sparsity or data similarity. Recovery of sparse signals from network coded data has been considered by several authors. Draper {\em et al.} \cite{compsensingFq} derive bounds on the error probability of a hypothetical decoder that recovers sparse sources from both noiseless and noisy measurements using the $l_{0}$-norm. The measurements are generated by randomly combining the source values in a finite field. Conditions on the number of measurements necessary for perfect recovery (with high probability) are provided and connections to the corresponding results in real field compressed sensing are established. Other works have as well investigated the sparse source recovery under the network coding framework \cite{NabaeeSSP2012,FeiziAllerton11,KattiAllerton07}. However, in these works the network coding operations are performed in the real field \cite{DeyNETCOD08}, which enables decoding using classical compressed sensing techniques \cite{CandesSPM08}, while, to the best of our knowledge, there is no practical scheme for recovering sparse data from incomplete network coded data in finite fields.

When the network coding operations are performed in a finite field, decoding from an incomplete set of network coded data becomes challenging and only a few works in the literature have proposed constructive solutions to this problem. The work in \cite{approxdecoding} presents an approximate decoding technique that exploits data similarity by matching the most correlated data in order to compensate for the missing network coded packets. The tradeoff between the data quantization and the finite field selection is studied, and the optimal field size for coding is established. In \cite{KiefferNetCod11}, robustness to lost network coded packets is achieved by combining network coding with multiple description coding (MDC). Redundancy is introduced through MDC, which results in graceful quality degradation as the number of missing network coded packets increases. The decoding is performed via mixed integer quadratic programming, which limits the choice of the finite field size to be equal to a prime number or to a power of a prime number.

Another class of decoding methods relies on the sum-product algorithm \cite{sumproduct}. In \cite{BarrosISWCS09}, the authors first build a statistical model for each network coded packet, which reflects the packet's path in the network. Then, they combine these models with the source statistics in order to obtain the overall statistical model of the system. This model is used for maximum a posteriori (MAP) decoding, which is solved using the sum-product algorithm on the corresponding factor graph. The main drawback of this approach is that it assumes global knowledge on the network topology, the paths traversed by the packets and the transition probabilities, which is not realistic in large scale dynamic networks. Linear network coding and the sum-product algorithm are also used in \cite{RajawatTWCOMM2012} for data gathering and decoding in wireless sensor networks. However, the proposed approach finds limited application in practical scenarios as it relies on a number of restrictive assumptions including the assumption that the random variables form a Markov Random Field which permits to factorize the joint pmf into low degree factors. Furthermore, the decoding error analysis is presented for specific types of joint pmfs that cannot be straightforwardly extended to general joint probability distributions.  The work in \cite{BassiEUSIPCO12} builds on the same ideas of using the sum product algorithm for decoding and employs a routing protocol to obtain a sparse coding matrix. Note that, even if our decoding algorithm also relies on the sum-product algorithm, it does not assume any knowledge of the network topology or the factorization of the joint pmf of the sources, contrarily to \cite{RajawatTWCOMM2012,BassiEUSIPCO12,BarrosISWCS09}. Our analysis of the decoding error probability is not tailored to a specific form of the joint source pmf, and the proposed decoding algorithm only requires the knowledge of the pairwise source correlation. In addition, our solution has significantly lower computational complexity compared to \cite{BarrosISWCS09,BassiEUSIPCO12,RajawatTWCOMM2012}.

From the theoretical perspective, the work by Ho {\em et al.} \cite{HoCorrelated} is among the earliest attempts to characterize the joint source and network coding problem for correlated sources. The communication scenario in \cite{HoCorrelated} consists of two arbitrarily correlated sources that communicate with multiple receivers over a general network where nodes perform randomized network coding. The authors provide upper bounds on the probability of decoding error as a function of network parameters, e.g., the min-cut capacities and the maximum source-receiver path length. The error exponents provided in \cite{HoCorrelated} generalize the Slepian-Wolf error exponents for linear coding \cite{CsiszarTIT82} and reduce to the latter in the special case where the network consists of separate links from each source to the receiver.

The limitations posed by the high complexity of the joint source-network coding problem have been partially addressed in several works about the practical design of the joint distributed source and network coding solutions. For example, a practical solution to the problem of communicating correlated information from two sources to multiple receivers over a network has been proposed in \cite{WuTIT09}. The source correlation is represented by a binary symmetric channel (BSC). The transmission is based on random linear network coding \cite{RandomizedNC03}, while the source compression is achieved by syndrome-based coding \cite{DISCUSTIT03}. Despite the linear operations in the network, the desirable code structure is preserved and enables low-complexity syndrome-based decoding. The high complexity of the joint decoding of source and network codes has also motivated research efforts in the direction of separating the source and the network coding \cite{EffrosDIMACS03}, \cite{Ramamoorthy06}.  However, it has been shown that the source and network coding cannot be separated in general multicast scenarios. The goal of the present paper is not to design a joint source and network coding scheme. We look at the complementary problem of recovering the source information from an incomplete set of network coded data. The key idea of our approach is to exploit the statistical properties of the sources, and in particular the source correlation, in order to provide approximate source reconstruction when the available network coded data is not sufficient for exact source recovery.


\section{Framework}
\label{sec:framework}

\begin{figure}[t]
	\begin{center}
		\includegraphics[width=0.65\textwidth]{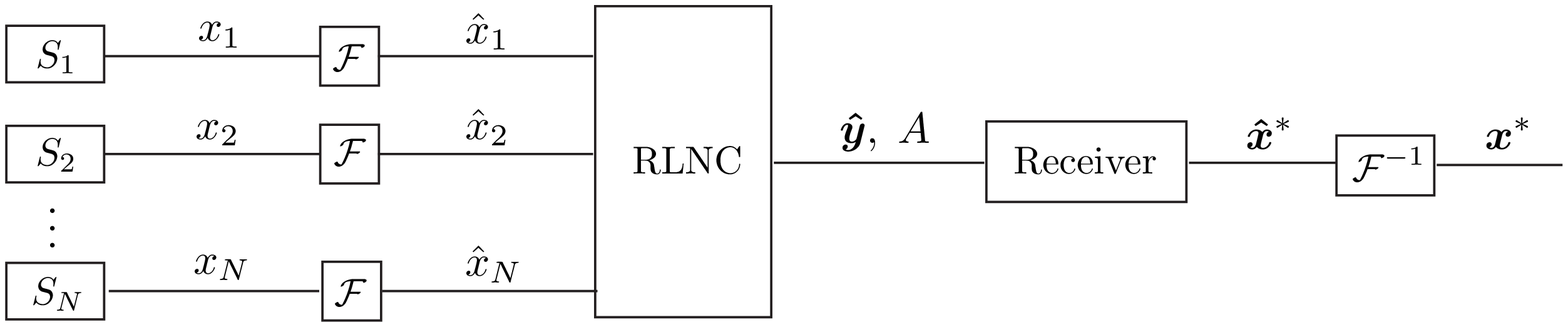}
	\end{center}
	\vspace{-0.9cm}
	\caption{Proposed network coding framework.}	
	\vspace{-0.3cm}
	\label{fig:framework}
\end{figure}

We consider the system setup illustrated in Fig.~\ref{fig:framework}. $S_1,S_2,\dots,S_N$ form a set of $N$ possibly statistically dependent sources. The sources produce a sequence of discrete symbols $x_1,x_2,\dots,x_N$ that are transmitted to a receiver. Without loss of generality, we assume that the source symbols belong to a finite alphabet $\mathcal{X}$ that  is a subset of the set of integer numbers $({\mathcal{X}\subset\mathbb{Z}})$. For continuous sources, this can be achieved by quantizing the output of the sources with some quantizer $Q$. We also assume that the source alphabet $\mathcal{X}$ is common for all the sources. The symbol $x_n$ produced by the $n$-th source can be regarded as a realization of a discrete random variable $X_n$. Thus, we represent the source $S_n$ by the random variable $X_n$ with probability mass function ${f_n(x):\mathcal{X}\rightarrow [0,1]}$. We also define the joint probability mass function (pmf) of the random vector $\bm X = (X_1, X_2, \dots, X_N)^T$ as ${f(\bm{x}):\mathcal{X}^{N}\rightarrow [0,1]}$.  

Prior to transmission, the source symbol $x_{n}$ is mapped to its corresponding representation  $\hat{x}_n$ in a Galois field of size $q$ through a bijective mapping ${\mathcal{F}: \mathcal{X}\rightarrow  \mathcal{F}(\mathcal{X})= \hat{\mathcal{X}} \subseteq \mathbb{F}_{q}}$, such that 
\begin{equation}
	{\hat{x}_n = \mathcal{F}[x_n] } \quad \mbox{and} \quad {x_n = \mathcal{F}^{-1}[\hat{x}_n], \; n = 1,\dots,N}
\label{eq:A1_ch3}
\end{equation}
The size $q$ of the field is chosen such that $|\mathcal{X}| \leq q$, where $|\mathcal{X}|$ denotes the cardinality of the source alphabet. 

We denote as $\hat{X}_{n}$ the random variable that represents the $n$-th source described in the Galois field domain. The marginal probability mass functions $\hat{f}_{n}(\hat{x}): \mathbb{F}_{q}\rightarrow [0,1]$ and the joint pmf $\hat{f}(\hat{\bm{x}}):\mathbb{F}_{q}^{N}\rightarrow [0,1]$ of the source symbol values represented in the Galois field $\mathbb{F}_{q}$ can be obtained from $f_{n}(x)$ and $f(\bm{x})$ respectively by setting

\begin{equation}
\hat{f}_{n}(\hat{x}) = 
\begin{cases}
	f_{n}(\mathcal{F}^{-1}[\hat{x}]),  &\mbox {if } \hat{x} \in \hat{\mathcal{X}}\\
	0, & \mbox{if } \hat{x} \in \mathbb{F}_{q} \backslash \hat{\mathcal{X}}
\end{cases} \quad \mbox{and} \quad \hat{f}(\hat{\bm{x}}) =
 \begin{cases}
	 f(\mathcal{F}^{-1}[\hat{\bm{x}}]), &\mbox {if } \hat{\bm{x}} \in \hat{\mathcal{X}}^{N}\\ 
	 0, & \mbox{if } \hat{\bm{x}} \in \mathbb{F}_{q}^{N} \backslash \hat{\mathcal{X}}^{N}
\end{cases}
\label{eq:A2_ch3}
\end{equation}

The source symbols given in $\mathbb{F}_{q}$ are transmitted to the receiver through intermediate network nodes. The intermediate network nodes perform random linear network coding and forward on their output links some random linear combinations of the symbols they receive. Thus, the $l$-th network coded symbol $\hat{y}_l$ that reaches the receiver can be written as 
\begin{equation}
	\hat{y}_l = \sum_{n=1}^N a_{ln}\hat{x}_{n}
\label{eq:A3_ch3}
\end{equation}
where $a_{ln}\in\mathbb{F}_{q}$ and all the arithmetic operations are performed in the Galois field $\mathbb{F}_{q}$. In this work, we focus on the case where the elements of the resulting matrix ${\bm A} = \{a_{ln}\}$ of global coding coefficients are uniformly distributed over $\mathbb{F}_q$.{\footnote{Such matrices arise in a variety of networking scenarios where data delivery/gathering is performed with the help of random linear network coding. One illustrative example is the data gathering in wireless sensor networks where data is collected along a spanning tree rooted at the gateway node. When the sensor nodes and the gateway node perform random linear network coding operations on the input symbols, the global coding coefficients of the network coded symbols that are sent from the gateway node to the receiver for further processing, are uniformly distributed over the Galois field.}}
The global coding coefficients are delivered along with the encoded symbols to the receiver to enable decoding. In practice, in order to reduce the overhead introduced by the coding coefficients, several source symbols can be concatenated in a single packet and the same coding coefficient can be used for all the symbols within a packet  \cite{approxdecoding}.

The receiver collects $L$ network coded symbols represented by a vector $\bm{\hat{y}} = (\hat{y}_1,\hat{y}_2,\dots,\hat{y}_L)^T$ such that 
\begin{equation}
	\bm{\hat{y}} = \bm{A}\hat{\bm{x}}
\label{eq:A4_ch3}
\end{equation} 

\noindent where $\bm{A}$ is a $L\times N$ matrix of global coding coefficients $\{a_{ln}\}$ and $\bm{\hat{x}} = (\hat{x}_1,\hat{x}_2,\dots,\hat{x}_N)^T$ is the vector of source symbols represented in $\mathbb{F}_{q}$. Given the vector of network coded symbols $\bm{\hat{y}}$ and the matrix of coding coefficients $\bm{A}$, the decoder provides an estimation $\bm{\hat{x}}^* $ of the vector $\bm{\hat{x}}$ which is subsequently mapped to the estimation $\bm{x}^* $ of the vector $\bm{x}= (x_1,x_2,\dots,x_N)^T$ of source symbols through the inverse mapping ${\mathcal{F}^{-1}: \hat{\mathcal{X}}\rightarrow \mathcal{X}}$, such that
 \begin{equation}
	 x_n^* = \mathcal{F}^{-1}[\hat{x}_n^*], \;n = 1,2,\dots,N
 \label{eq:A5_ch3}
 \end{equation}
where $\hat{x}_n^* \in\hat{\mathcal{X}} \subseteq\mathbb{F}_{q}$ and ${x}_n^* \in \mathcal{X}$.

If the rank of the matrix of coding coefficients $\bm{A}$ is equal to the number of source symbols $N$, then the matrix $\bm{A}$ is invertible and the source symbols can be perfectly decoded, {\em i.e.}, $x^{*}_{n} = x_{n}, \; \forall n$. However, if the number of linearly independent network coded symbols is smaller than $N$, conventional decoding methods, such as Gaussian elimination, can generally not recover any source symbol from Eq.~\eqref{eq:A4_ch3}. In this paper we exactly focus on this specific case, where only incomplete information is available at the decoder. Our first objective is to provide a theoretical characterization of the performance of the proposed framework in terms of decoding error probability under {\em maximum a posteriori} decoding. We then focus on a particular class of statistically dependent sources that present pairwise linear correlation and we design a practical decoding algorithm that permits approximate reconstruction of the source symbols with the help of some prior knowledge about the sources.


\section{Performance analysis under MAP decoding}
\label{sec:error}

In this section, we analyze the performance of the network coding system presented in Section \ref{sec:framework}. In particular, we derive an upper bound on the error probability of a MAP (maximum a posteriori) decoder. A MAP decoder selects the sequence of source symbols ${\hat{\bm x}}^{*}$ that maximizes the a posteriori probability given the observation $\hat{{\bm y}}$ of $L$ network coded symbols and the matrix $\bm{A}$ of coding coefficients. The maximum a posteriori decoding rule is optimal in the sense that it minimizes the probability of decoding error for a given set of source sequences and a given encoding procedure  \cite{GallagerITRCbook}. Though exact MAP decoding for general linear codes is known to be NP-complete and its implementation complexity is prohibitively high especially for long sequences \cite{InferenceMackay}, it can still give useful insights into the performance limits of our framework. In order to establish the upper bound on the decoding error probability, we follow the development of \cite{SourceCodingGallager}, where the author studies the problem of source coding with side information. Indeed, the framework that we examine bears some resemblance to the source coding problem and can also be viewed as an instance of distributed in-network compression. 

Formally, the MAP decoding rule can be written as
\begin{equation}
\begin{split}
	{\hat{\bm x}}^{*} &= \argmax_{\hat{\bm x} \in \hat{\mathcal{X}}^{N}}p(\hat{\bm x}|{\hat{{\bm y}}},\bm{A})  
	= \argmax_{\hat{\bm x} \in \hat{\mathcal{X}}^{N}}\frac{p({\hat{{\bm y}}}|\hat{\bm x},\bm{A}) p(\hat{\bm x}|\bm{A})}{p(\hat{\bm{y}}|\bm{A})} 
	\label{eq:D1_ch3}
	\end{split}
	\end{equation}
The probability $p(\hat{\bm{y}}|\hat{\bm{x}},\bm{A})$ of receiving a vector $\hat{\bm{y}}$ of network coded symbols conditioned on the event that a vector $\hat{\bm{x}}$ of source symbols has been transmitted and encoded with the coding matrix $\bm{A}$, is equal to 1 if $\bm{A}\hat{\bm x} = \hat{\bm y}$, and equal to 0 otherwise. Given the fact that $p(\hat{\bm{y}}|\bm{A})$ does not depend on $\hat{\bm{x}}$ and that the vector $\hat{\bm{x}}$ and the matrix $\bm{A}$ are statistically independent, {\em i.e.}, $p(\hat{\bm{x}}|\bm{A}) = p(\hat{\bm{x}}) = \hat{f}(\hat{\bm{x}})$, the decoding rule presented in Eq.~\eqref{eq:D1_ch3} can be further simplified as 
\begin{equation}
	{\hat{\bm x}}^{*} =\argmax_{\substack{\hat{\bm x} \in \hat{\mathcal{X}}^{N}}} \mathbbm{1}_{\{\bm{A}, \hat{\bm y}\}}(\hat{\bm x})\hat{f}(\hat{\bm x})
	\label{eq:D2_ch3}
\end{equation}
where $\mathbbm{1}_{\{\bm{A}, \hat{\bm y}\}}(\hat{\bm x})$ is an indicator function defined as
\begin{equation}
	\mathbbm{1}_{\{\bm{A}, \hat{\bm y}\}}(\hat{\bm x}) = 
	\begin{cases}
		1, & \text{ if } \bm{A}\hat{\bm x} = \hat{\bm y} \\
		0, & \text{ otherwise}
	\end{cases}
\label{eq:D3_ch3}
\end{equation}
\begin{proposition}
For a coding matrix $\bm{A}$ with i.i.d. entries uniformly distributed over a Galois field $\mathbb{F}_{q}$, the probability of error $P_{e}$ under the MAP decoding rule (Eq.~\eqref{eq:D2_ch3}) is upper bounded by
\begin{equation}
		P_{e} \leq \min_{0\leq \rho \leq 1} 2^{-\rho L\log_2 q +\rho H_\rho (\bm X) - D_{KL} (f_\rho(\bm x)|| f(\bm x))} \label{eq:D4_1_ch3} 
\end{equation}
where $L$ is the number of network coded symbols available at the decoder, $f(\bm x)$ is the joint pmf of the sources and $\rho \in [0,1]$ is a scalar. $f_\rho (\bm x)$ and $H_\rho (\bm X)$ are respectively the tilted distribution and its entropy, defined as 
\begin{subequations}
\begin{align}
	f_{\rho}(\bm{x})  &= \frac{f(\bm{x})^{\frac{1}{1+\rho}}}{\sum\limits_{\bm{x}\in \mathcal{X}^{N}}f(\bm{x})^{\frac{1}{1+\rho}}}
	\label{eq:D18_ch3} \\
\intertext{and} \nonumber \\[-6ex]
	H_{\rho}( \bm{X} ) &= - \sum\limits_{\bm{x} \in \mathcal{X}^{N}} f_{\rho}(\bm{x})\log_{2}f_{\rho}(\bm{x})
\label{eq:D19_ch3}
\end{align}
\end{subequations}
respectively, and $D_{KL}(f_\rho(\bm x) || f(\bm x))$ is the Kullback-Leibler divergence of $f(\bm x)$ from $f_\rho(\bm x)$.
\label{prop:1}
\end{proposition}
\noindent The proof of Proposition \ref{prop:1} is given in Appendix \ref{app:1}.

Let us now further investigate the behavior of the upper bound on the probability of error with respect to the parameters $L$, $q$ and the joint probability mass function $f(\bm{x})$, and determine the value of $\rho$ that minimizes the expression of the upper bound. To do so, we first show that the entropy $H_\rho (\bm X)$ is a non-decreasing function of $\rho$ by using the following proposition: 
\begin{proposition}
The function $E(\rho)$ 
\begin{subequations}
\begin{align}
E(\rho) =&-\rho L\log_2 q +\rho H_\rho (\bm X) - D_{KL} (f_\rho(\bm x)|| f(\bm x)) \\
 =&-\rho L \log_{2} q+ (1+\rho)\log_{2}\Big{[}\sum_{{\bm x}\in \mathcal{X}^{N}}f(\bm x)^{\frac{1}{1+\rho}}\Big{]}
	\label{eq:D12_ch3}
\end{align}
\end{subequations}
 is a convex function of $\rho$ for $\rho \geq 0$ with strict convexity unless the random vector $\bm X = (X_{1}, X_{2},\dots , X_{N})$ is uniformly distributed, \em{i.e.}, $f(\bm{x}) = \frac{1}{|\mathcal{X}^{N}|}$, $\forall \bm{x} \in \mathcal{X}^{N}$.
\label{prop:2}
\end{proposition}
\noindent The proof of Proposition \ref{prop:2} is given in Appendix \ref{app:2}. Since $E(\rho)$ is the base-2 logarithm of the upper bound in Eq. \eqref{eq:D4_1_ch3}, the value of $\rho$ that minimized $E(\rho)$, also minimizes the value of the upper bound. The first partial derivative of $E(\rho)$ is equal to 
\begin{equation}	
		\frac{\partial E(\rho)}{\partial \rho} = -L \log_{2}q  + H_{\rho}( \bm{X} )
\label{eq:D17_ch3}
\end{equation}
Since $E(\rho)$ is convex, its second partial derivative with respect to $\rho$ is non negative. We have
\begin{equation}
		\frac{\partial ^{2} E(\rho)} {\partial \rho^{2}} = \frac{\partial H_{\rho}(\bm{X})}{\partial \rho} \geq 0 
		\label{eq:D20_ch3}
\end{equation}
From Eq.~\eqref{eq:D20_ch3} it follows that $H_{\rho}(\bm{x})$ is a non-decreasing function of $\rho$, for $0 \leq \rho \leq 1$. We can, therefore, identify the following three cases regarding the upper bound on the probability of error:
\begin{itemize}
\item If $ L \log_{2} q < H_{\rho}(\bm{X})|_{\rho = 0}= H(\bm{X})$, then $ \frac{\partial E(\rho)}{\partial \rho} > 0$ for all $0 \leq \rho \leq 1$. Given that $E(\rho)$ is convex and that its first partial derivative does not change sign and is positive in the interval $0 \leq \rho \leq 1$, the value of $\rho$ that minimizes the upper bound is $\rho = 0$. Thus, we obtain the trivial bound $$P_{e} \leq 1$$
\item If $L \log_{2} q > H_{\rho}(\bm{X})|_{\rho = 1} $, then $ \frac{\partial E(\rho)}{\partial \rho}  <  0$ for all $0 \leq \rho \leq 1$. Given that $E(\rho)$ is convex and that its first partial derivative does not change sign and is negative in the interval $0 \leq \rho \leq 1$, the value of $\rho$ that minimizes the upper bound is $\rho = 1$. Thus, we obtain the  bound $$P_{e} \leq 2^{-L\log_2 q + H_\rho (\bm X) - D_{KL} (f_\rho(\bm x)|| f(\bm x))}|_{\rho = 1}$$
\item Finally, if $H(\bm{X}) \leq L \log_{2} q \leq H_{\rho}(\bm{X})|_{\rho = 1}$, the value of $\rho$ that minimizes the expression given in Eq.~\eqref{eq:D12_ch3} and, therefore, the upper bound on the error probability in Eq.~\eqref{eq:D4_1_ch3}, can be obtained analytically by setting the first partial derivative of $E(\rho)$ equal to $0$
\begin{equation}
	 L\log_{2}q = H_{\rho}({\bm X})
\label{eq:D21_ch3}
\end{equation}
In this case, the upper bound on the probability of error is given by 
$$P_{e} \leq 2^{- D_{KL} (f_{\rho^{*}}(\bm x)|| f(\bm x))}$$
where $\rho^{*}$ is the solution of Eq.~\eqref{eq:D21_ch3}.
\end{itemize}

We now illustrate the behavior of the upper bound on the error probability for a particular instance of the network coding framework presented in Section \ref{sec:framework}. We consider a scenario with $N$ statistically dependent sources that generate discrete symbols from the alphabet $\mathcal{X}$. We assume that the joint probability mass function of the sources, $f(\bm{x})$, can be factorized as follows
\begin{equation}
	f(\bm{x})  = \prod\limits_{n=1}^{N}f(x_{i}|x_{i-1})
	\label{eq:D22_ch3}
\end{equation}
and we set the conditional probability mass functions to be equal to 
\begin{equation}
	f(x_{i}|x_{i-1} )= \frac{1}{K}\frac{1-p}{1+p}\;p^{\;|x_{i}-x_{i-1}|}, \quad p \in (0,1)
	\label{eq:D23_ch3}
\end{equation}
\noindent where $K$ is a normalization constant such that $\sum\limits_{x_{i} = 0}^{q-1}f(x_{i}|x_{i-1} ) = 1$. For a given value of $x_{i-1}$, the distribution in Eq.~\eqref{eq:D23_ch3} represents a shifted and truncated discrete Laplacian distribution \cite{discreteLaplace}. The smaller is the value of the parameter $p$ in Eq.~\eqref{eq:D23_ch3}, the higher is the correlation between the variables $X_{i-1}$ and $X_{i}$. We also assume that $\mathcal{F}(\mathcal{X}) \equiv \mathbb{F}_{q}$ for some value of $q$ and that $X_{1} $ is uniformly distributed over $\mathcal{X}$. The source symbols are transmitted to the receiver through an overlay network, where intermediate network nodes perform randomized linear network coding. The receiver collects a set of $L$ network coded symbols and reconstructs the source data by applying the MAP decoding rule given in Eq.~\eqref{eq:D2_ch3}.

\begin{figure}[t]
	\begin{center}
		\subfloat[]{\includegraphics[width = 0.45 \textwidth]{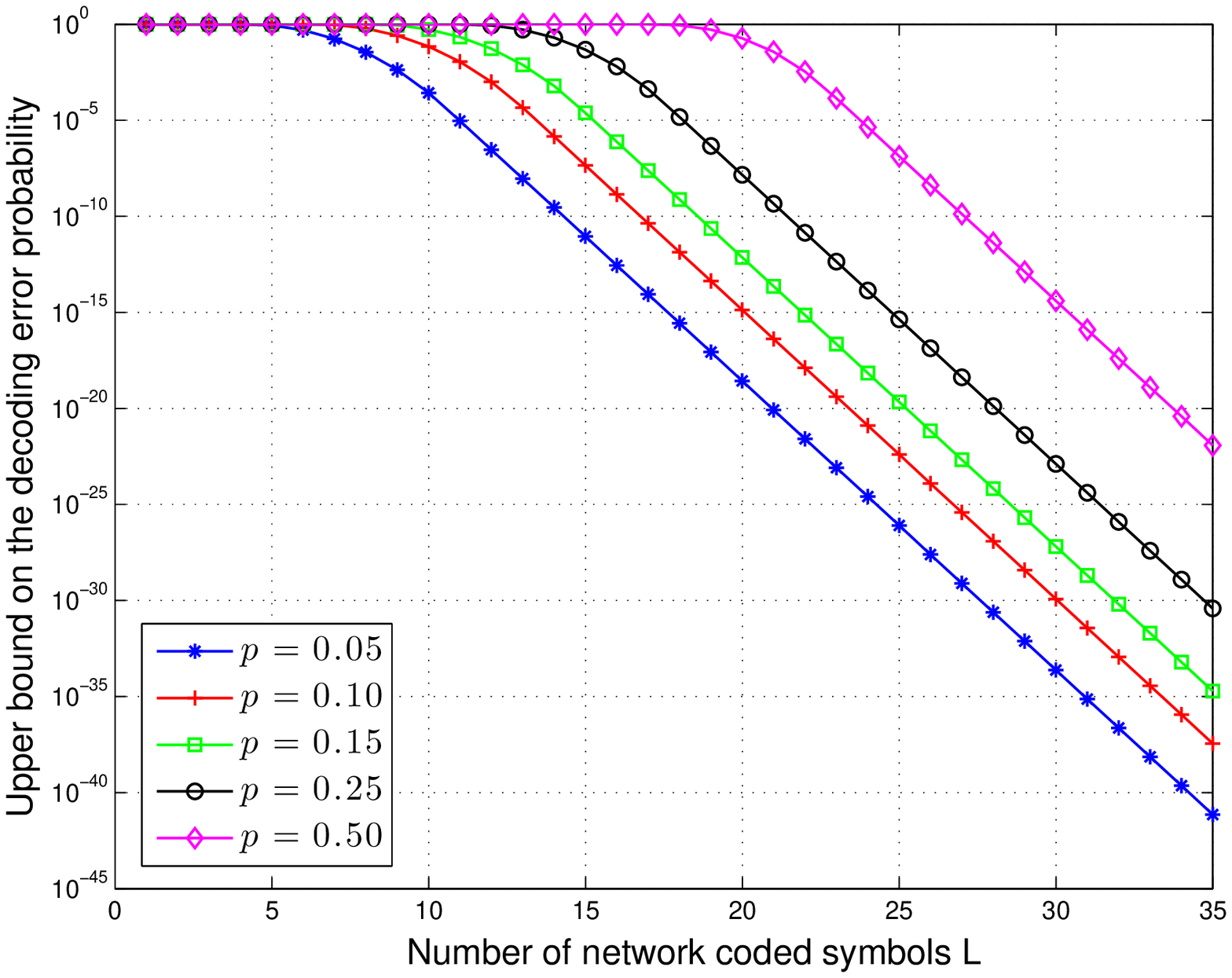}}
		\subfloat[]{\includegraphics[width = 0.45 \textwidth]{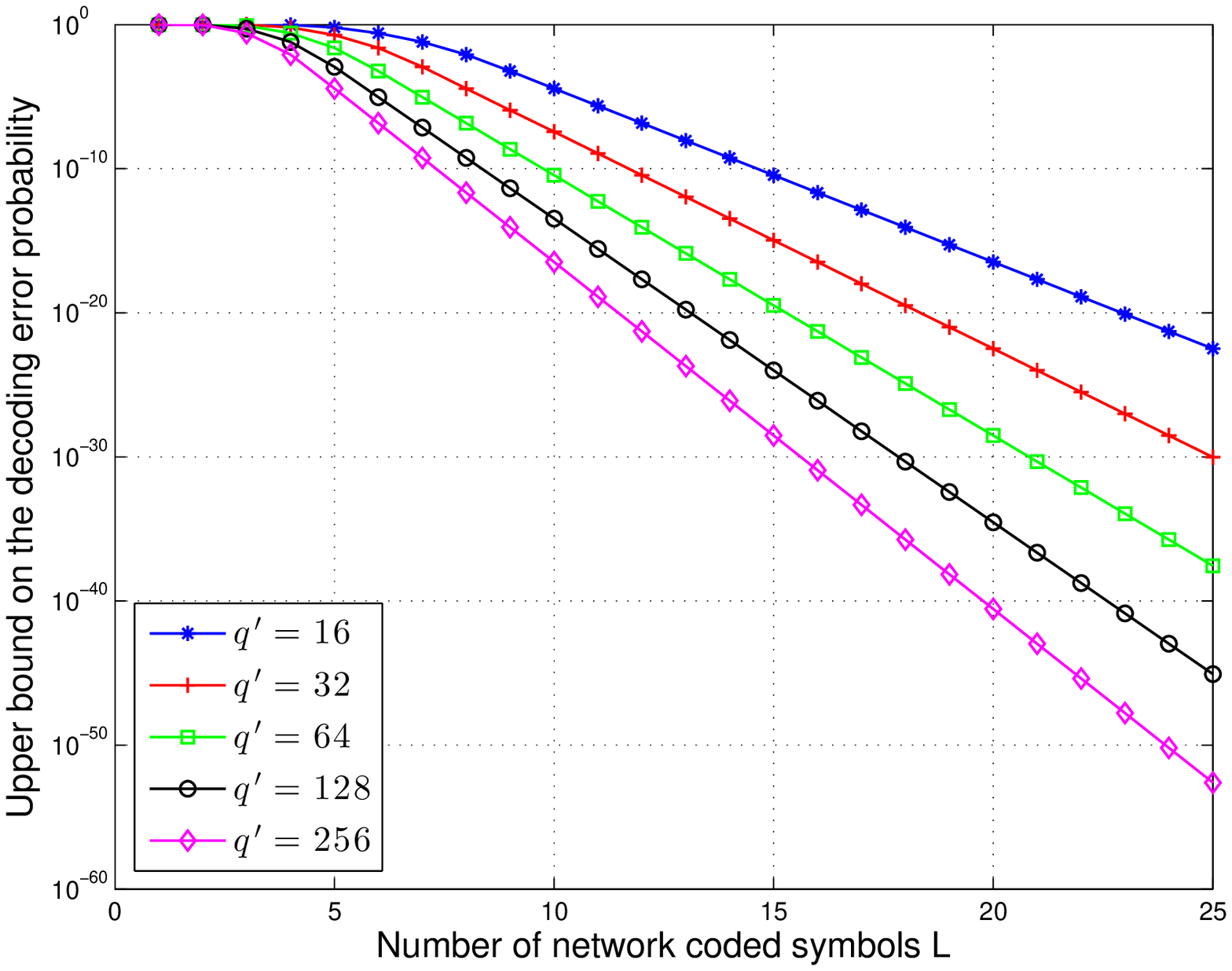}}
	\end{center}
	\vspace{-0.4cm}
	\caption{Upper bound on the decoding error probability versus the number of network coded symbols $L$ available at the receiver (a) for $N = 30$ sources, $p =\{ 0.05, 0.10, 0.15, 0.25, 0.50\}$ and $q = 32$, and (b) for $N = 20$ sources, $p = 0.05$ and $q^{\prime} = \{16, 32, 64, 128, 256\}$. }
	\vspace{-0.4cm}
	\label{fig:bound}
\end{figure}

Fig.~\ref{fig:bound}(a) illustrates the upper bound on the decoding error probability for different values of the parameter $p$ with respect to the number of network coded symbols $L$ available at the decoder. As expected, higher correlation ({\em i.e.}, lower value of $p$) leads to a lower probability of error and, thus, to a better performance of the MAP decoder. Note that the error probability decreases as the number of network coded symbols increases. However, even for $L \geq N$ the decoding error probability is non zero. This is due to the fact that the entries of the coding matrix $\bm{A}$ are uniformly distributed and the probability that $\mbox{rank}(\bm{A}) < N$ for $L \geq N$ is non zero, though this probability becomes negligible as the number of network coded symbols exceeds the number of original source symbols. 

In Fig.~\ref{fig:bound}(b), we show the evolution of the bound on the decoding error probability with the size of the Galois field that is used to represent the source data and perform the network coding operations. Here we assume that the cardinality of the source alphabet is equal to $q = 16$, while the size of the finite field used for data representation and network coding operations is equal to $q^{\prime}$ with $q^{\prime} \geq q$. From Fig.~\ref{fig:bound}(b) we can see that for a fixed number of network coded symbols, the bound on the probability of error decreases as the size of the Galois field increases, which indicates that by employing larger finite fields better performance can be achieved at the decoder for the same number of network coded symbols. This is due to the fact that by using larger Galois fields for the representation of the source data, we introduce some redundancy that assists the decoder to eliminate candidate solutions that are not valid according to the data model. Moreover, the probability of generating linearly dependent network coded symbols is lower in larger fields, thus every network coded symbol that arrives at the decoder brings novel information with higher probability and limits the solution space. However, this performance improvement comes at the cost of a larger number of bits that need to be transmitted to the receiver.

\begin{figure}[t]
	\begin{center}
		\subfloat[]{\includegraphics[width = 0.45 \textwidth]{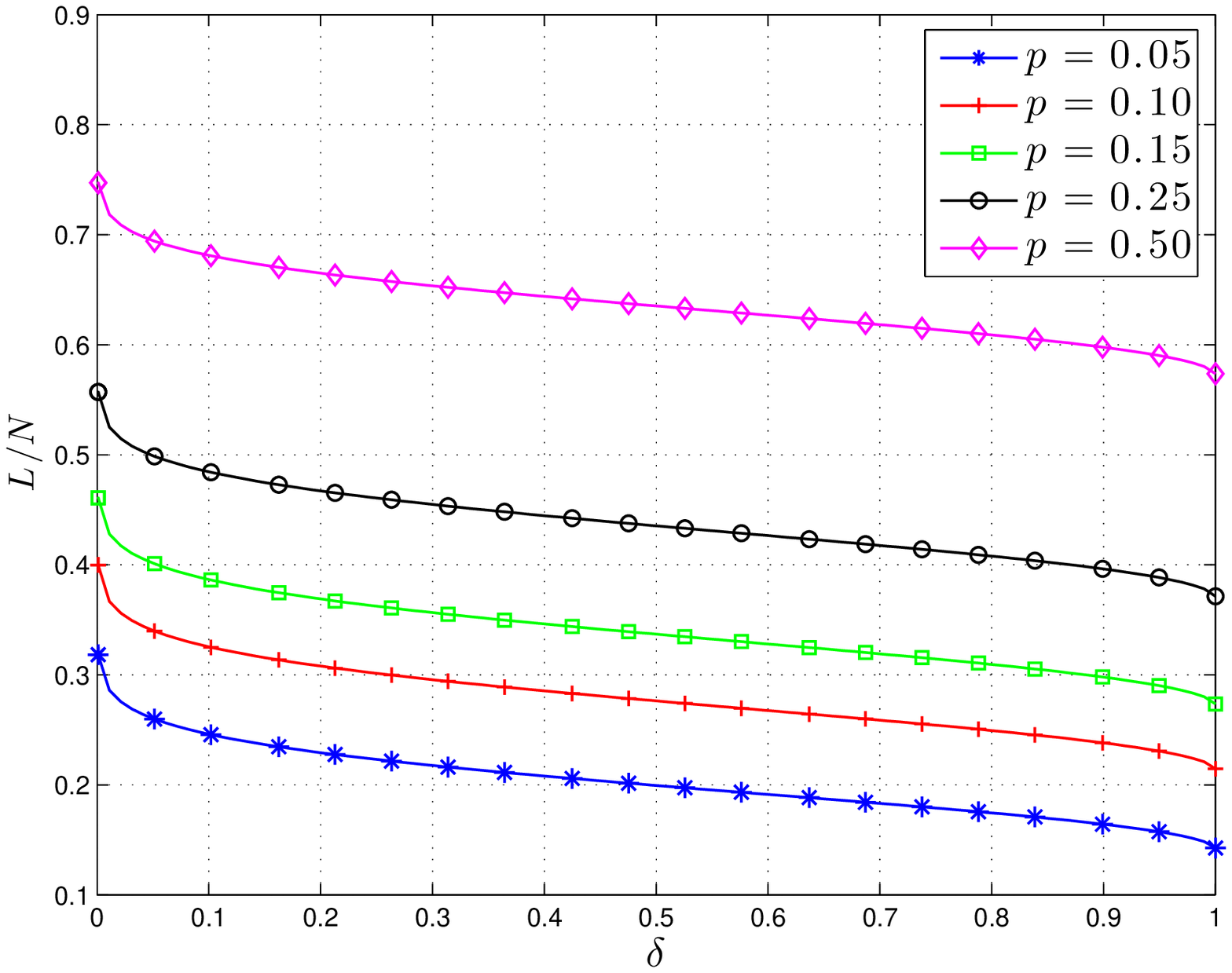}}
		\subfloat[]{\includegraphics[width = 0.45 \textwidth]{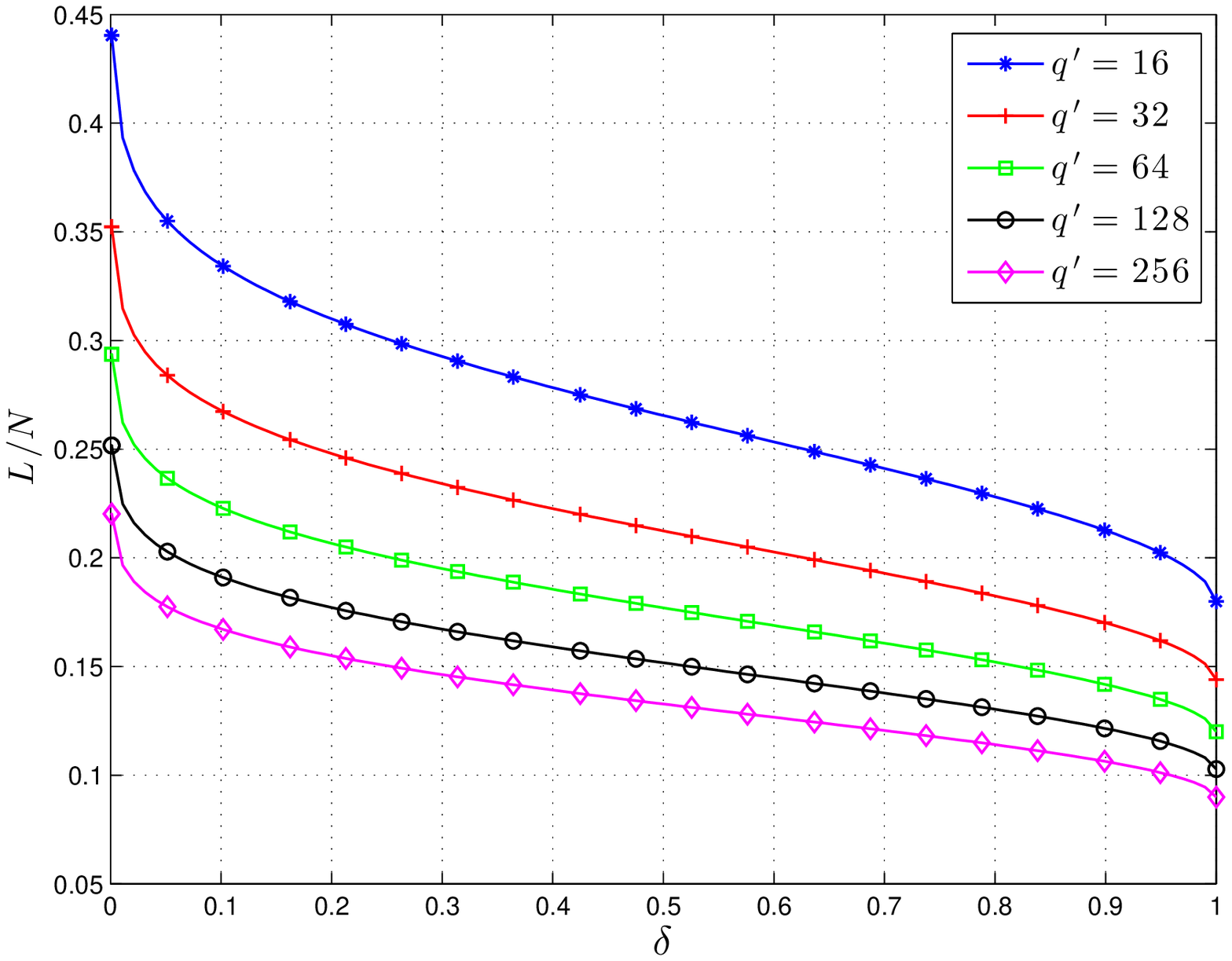}}
	\end{center}
	\vspace{-0.4cm}
	\caption{Lower bound on the number of network coded symbols $L$ versus the decoding error probability (a) for $N = 30$ sources, $p =\{ 0.05, 0.10, 0.15, 0.25, 0.50\}$ and $q = 32$, and (b) for $N = 20$ sources, $p = 0.05$ and $q^{\prime} = \{16, 32, 64, 128, 256\}$. }
	\vspace{-0.4cm}
	\label{fig:lowerbound}
\end{figure}

We finally derive the sufficient conditions on the number of network coded symbols $L$ required at the receiver to achieve a decoding error probability below a certain value $\delta$. 
 
 \begin{proposition}
For a coding matrix $\bm{A}$ with i.i.d. entries uniformly distributed over a Galois field $\mathbb{F}_{q}$, one can achieve a probability of error $P_{e} \leq\delta$ under the MAP decoding rule (Eq.~\eqref{eq:D2_ch3}), for $\delta \in (0,1)$ and for any $\rho \in (0,1]$, if the following condition on the number of network coded symbols $L$ is satisfied:
\begin{equation}
\frac{L}{N} \geq -\frac{\log_2 \delta}{\rho N \log_2 q} + \frac{H_\rho(\bm X)}{N \log_2 q} - \frac{D_{KL}(f_\rho(\bm x)||f(\bm x))}{\rho N \log_2 q}
\label{eq:L/N}
\end{equation}
 \label{prop:3}
 \end{proposition}
 
 The lower bound in Eq. \eqref{eq:L/N} follows immediately from Eq. \eqref{eq:D4_1_ch3} by upper bounding the right side part by $\delta \in (0,1)$ and taking the base-2 logarithm of both sides of the resulting inequality.  
 
 Fig.~\ref{fig:lowerbound} illustrates the lower bound on the number of network coded symbols $L$ required at the receiver to achieve a decoding error probability below $\delta$ for the set of sources with joint probability mass function given by Eqs~\eqref{eq:D22_ch3} and \eqref{eq:D23_ch3}. Fig.~\ref{fig:lowerbound}(a) depicts the minimum number of network coded symbols $L$ (normalized over the number of sources $N$) required at the decoder for different values of the parameter $p$. We can observe that for sources with higher correlation ({\em i.e.,} lower value of $p$), a smaller number of network coded symbols is required to achieve a decoding error probability below the same value $\delta$. Fig.~\ref{fig:lowerbound}(b) shows the evolution of the lower bound on the number of network coded symbols $L$ with respect to the target decoding error probability $\delta$ for different values of the size of the Galois field that is used to represent the source data and perform the network coding operations. Fig.~\ref{fig:lowerbound}(b) shows that, when more bits are used to represent the source symbols, the minimum number of network coded symbols needed to stay below a certain decoding error probability reduces due to the redundancy introduced in the data representation. 
 
The theoretical analysis presented in this section is general and is valid for any set of sources with an arbitrary joint statistical model expressed in terms of the joint pmf $f(\bm{x})$. It permits to understand the influence of the different design parameters on the decoding performance. However, the exact solution of the MAP decoding problem described in Eq.~\eqref{eq:D2_ch3} requires the enumeration of all possible configurations of $N$ random variables that satisfy the network coding constraints. This becomes intractable as the number of sources and the size of the Galois field increase. In the next section, we propose a practical decoder that provides suboptimal, yet effective decoding performance for a class of statistically dependent sources that present pairwise linear correlation.


\section{Iterative decoding via message passing}
\label{sec:decoding}

We now design a low-complexity iterative message passing algorithm for decoding of network coded correlated data. The algorithm provides an estimation $\hat{\bm{x}}^{*}$ of the transmitted sequence $\hat{\bm{x}}$ of source values by jointly considering the network coding constraints and the prior statistical information about the source values. The proposed decoding algorithm is a variant of the standard Belief Propagation (BP) algorithm \cite{probreasoning,LDPCqMacKay} with source priors that are directly incorporated in the messages exchanged over the factor graph. In the next subsections, we describe in detail the correlation model, the factor graph representation that captures the network coding operations, and the complete approximate decoding algorithm based on BP.


\subsection{Statistical source priors}
\label{sec:correlationmodel_ch3}

We focus on the case where the sources are linearly correlated and we assume, without loss of generality, that the correlation coefficient $\rho_{ij}$ between the variables $X_{i}$ and $X_{j}$, $i \neq j$ is non negative ($\rho_{ij}\geq 0$), {\em i.e.}, the sources are positively correlated. In this case, the pairwise statistical relationships between the sources can be efficiently represented as a correlation noise. Hence, for every pair of sources $S_{i}$ and $S_{j}$, $i,j \in \{1,2,\dots,N\}$ with $i\neq j$, we define the discrete random variable $W_{m}$ such that
\begin{equation}
	W_{m} = X_{i} -X_{j} , \quad m = 1,2,\dots, M
	\label{eq:B1_ch3}
\end{equation}
where $M $ is the number of correlated pairs of sources. We denote as ${g_{m}(w): \mathcal{W} \rightarrow [0,1]}$ the probability mass function of the random variable $W_{m}$, where $\mathcal{W}$ is the set of all possible values of the difference of random variables $X_{i} - X_{j}$. Note that for a pair of sources with a negative correlation coefficient $\rho_{ij}$, a similar approach can be used by defining the correlation noise between the random variables $X_{i}$ and $-X_{j}$.

The number of correlated pairs of variables can be as high as $N(N-1)/2$ in the case where every source is correlated with each of the $N-1$ remaining sources. In practice, the source sequence is usually characterized by some local structure, for example localized spatial or temporal correlation. This localization of statistical dependencies limits the number of source pairs with significant correlation coefficients. The correlated source pairs can be identified by observing the local interactions in the underlying physical process captured by the sources. 

\begin{figure*}[t!]
	\begin{center}
			\subfloat[Sensor network: the data sources are spatially correlated.]{\label{fig:sensorexample}\includegraphics[width=0.45\textwidth]{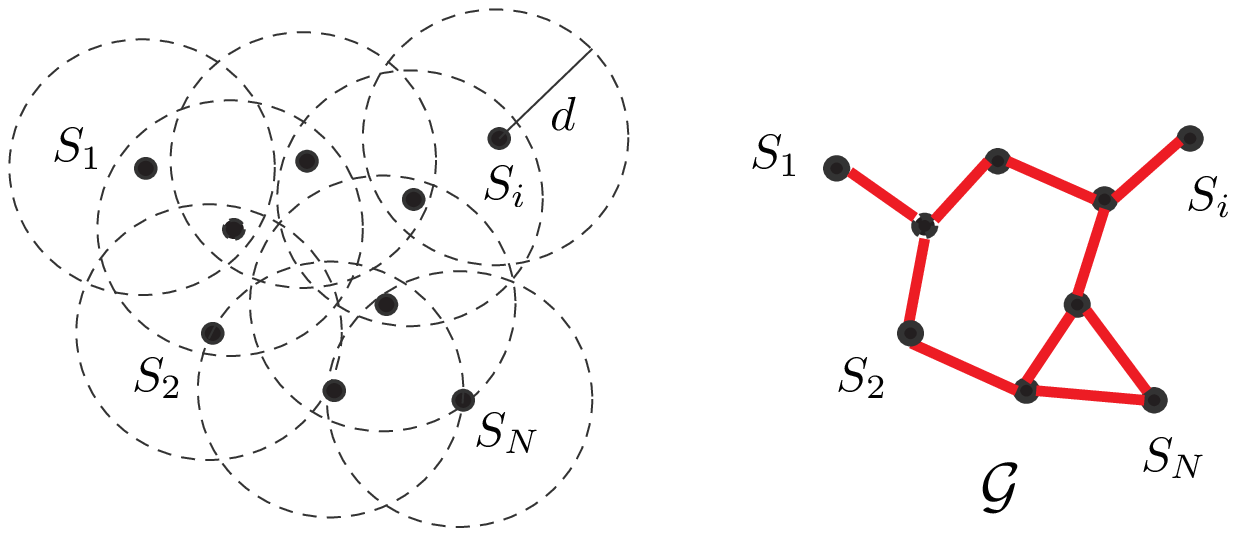}} \quad \quad \quad
			\subfloat[Image sequences: the data sources are temporally correlated.]{\label{fig:videoexample}\includegraphics[width=0.45\textwidth]{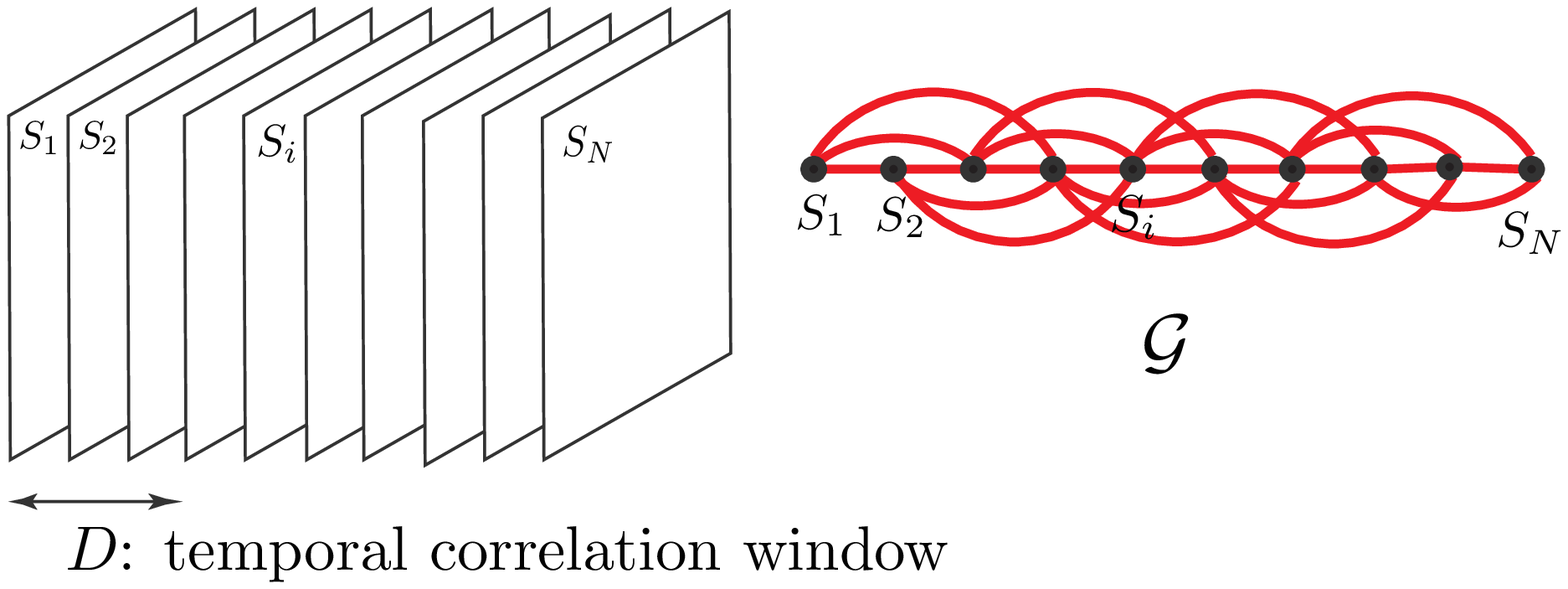}}
	\end{center}
	\vspace{-0.2cm}
	\caption{Examples of applications with correlated data sources and the corresponding undirected graphical models that represent the statistical relationships among the sources. \label{fig:correlationexample}}
	\vspace{-0.4cm}
\end{figure*}

It is convenient to represent the correlation structure by an undirected graph $\mathcal{G} = {(\mathcal{V}, \mathcal{E})}$ where each vertex corresponds to a source and two vertices $i$ and $j$ are connected with an undirected edge if the sources $S_{i}$ and $S_{j}$ are correlated, {\em i.e.}, $\rho _{ij}  \neq 0$. Two examples of applications with correlated data sources and the corresponding undirected graphs that capture the correlation between source pairs are illustrated in Fig.~\ref{fig:correlationexample}. In Fig.~\ref{fig:correlationexample}(a), we present a network of sensors. In this scenario, it is reasonable to assume that the value of the signal acquired by the sensor $S_{i}$ is correlated with the values generated by the sensors positioned within the distance $d$, with the correlation coefficient $\rho_{ij}$ decaying as the distance between the sensors increases. Fig.~\ref{fig:correlationexample}(b) shows a sequence of images captured at different time instants, where the pixel values in the $i$-th image are correlated with the pixel values of every other image that falls within the same time window of width $D$.


\subsection{Factor graph representation}
\label{sec:graph_ch3}

\begin{figure}[t]
	\begin{center}
			\includegraphics[width=0.5\textwidth]{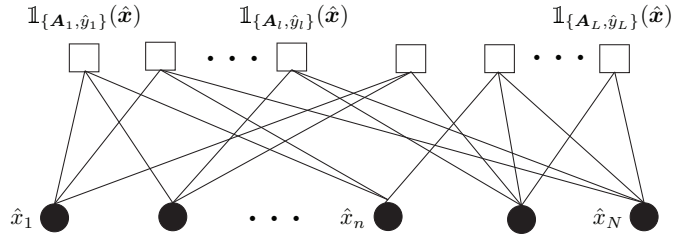}
	\end{center}
	\vspace{-0.7cm}
	\caption{ Factor graph. \label{fig:factorgraph}}
	\vspace{-0.3cm}
\end{figure}

The constraints imposed by network coding operations can be described by a factor graph as the one illustrated in Fig.~\ref{fig:factorgraph}. This bipartite graph consists of $N$ variable nodes and $L$ check nodes that form the basis of the message passing algorithm used for decoding. The variable nodes represent the source symbols, which are the unknowns of our problem. The check nodes represent the network coding constraints imposed on the source symbols. The $l$-th check node corresponds to the $l$-th network coded symbol and is connected to all the variable nodes that participate in that particular network coding combination. The check node is associated with the indicator function $\mathbbm{1}_{\{\bm{A}_{l}, \hat{ y}_{l}\}}(\hat{\bm x})$, which is defined as 
\begin{equation}
	\mathbbm{1}_{\{\bm{A}_{l}, \hat{ y}_{l}\}}(\hat{\bm x}) = 
	\begin{cases}
		1, & \text{ if } \bm{A}_{l}\hat{\bm x} = \hat{y}_{l} \\
		0, & \text{ otherwise}
	\end{cases}
\label{eq:B2_ch3}
\end{equation}
The indicator function $\mathbbm{1}_{\{\bm{A}_{l}, \hat{ y}_{l}\}}(\hat{\bm x})$ takes the value 1 when a certain configuration $\hat{\bm x}$ satisfies the equation defined by the $l$-th row of the coding matrix $\bm{A}$ and the value of  the $l$-th network coded symbol $\hat{y}_{l}$, and the value 0 otherwise. The degree of the $l$-th check node is equal to the number of non zero network coding coefficients in the $l$-th row of the coding matrix $\bm{A}$, while the total number of check nodes is equal to the number $L$ of network coded symbols available at the decoder.

One of the parameters that influence the computational complexity and the convergence speed of the message passing algorithms is the degree of the check nodes. Recall that, in our setting, the elements of the coding matrix $\bm A$ are uniformly distributed over $\mathbb{F}_q$. That means that the matrix $\bm A$ is dense and every network coded symbol is a linear combination of nearly all the source symbols. Thus, the degree of the check nodes that correspond to the network coding constraints is nearly $N$. In order to reduce the degree of the check nodes, we preprocess the matrix of coding coefficients $\bm{A}$ before constructing the factor graph. This preprocessing consists in performing elementary row operations on the original matrix $\bm{A}$ in order to eliminate some of the coding coefficients. In particular, we first perform Gaussian elimination on the matrix $\bm A$. This results in a $L^\prime \times N$ matrix ($L^\prime \leq \min (L,N)$) with zeros below the main diagonal. We then eliminate $L^\prime-(j+1)$ elements from the $(N-j)$-th column for $j = 0, 1,\dots, L^\prime-2$ by performing elementary row operations in such a way that the elements below the main diagonal remain null. The vector of network coded symbols $\hat{\bm y}$ is subject to the same elementary row operations yielding $\hat{\bm y}^\prime$, such that the systems of equations $\hat{\bm{y}} = \bm{A}\hat{\bm{x}}$ and $\bm{\hat{y}}^\prime = \bm{A}^\prime\hat{\bm{x}}$ have the same set of solutions. Besides reducing the density, this preprocessing of the original coding matrix $\bm{A}$ helps to identify and eliminate the non-innovative network coded symbols.

Note that, unlike the factor graphs used in the standard sum-product algorithm \cite{sumproduct}, the factor graph in Fig.~\ref{fig:factorgraph} does not correspond to a factorization of some probability distribution function. It serves as a graphical model to capture the constraints imposed on the source symbols by the network coding operations.


\subsection{Message passing algorithm}
\label{sec:algo_ch3}

We now present our message passing algorithm for decoding of network coded correlated data that operates on the factor graph presented in Section \ref{sec:graph_ch3}. We begin by introducing the notation used throughout this section. 
We first define the following sets: 
 
 \begin{itemize}
 
 \item $\mathcal{L}(n) = \{l: a_{ln}^\prime \neq 0\}$ as the set of all the check nodes that are neighbours of the $n$-th variable node, and 
  \item $\mathcal{N}(l) = \{n: a_{ln}^\prime \neq 0\}$ as the set of all the variable nodes that are neighbours of the $l$-th check node,
 \end{itemize}
 where the node $i$ is called a \textit{neighbour} of the node $j$ if there exists an edge between nodes $i$ and $j$ in the factor graph. The messages that are exchanged over the factor graph are defined and interpreted as follows:
 \begin{itemize}
 \item $q_{nl}(a)$, $a\in \mathbb{F}_{q}$, denote the messages sent from the $n$-th variable node to the $l$-th check node. These messages represent the belief of the $n$-th variable node that source symbol $\hat{x}_n$ has the value $a$, given the information obtained from all the neighbours other than the $l$-th check node, 
 \item $r_{ln}(a)$, $a\in \mathbb{F}_{q}$, denote the messages sent from the  $l$-th check node to the $n$-th variable node. These messages represent the belief of the $l$-th check node that the source symbol $\hat{x}_n$ has value $a$, given the messages from all the variable nodes other than the $n$-th variable node.
 \end{itemize}

Equipped with this notation, we describe now the decoding process in more details. The algorithm takes as input the matrix $\bm{A}^{\prime}$, the vector $\hat{\bm{y}}^{\prime}$, the adjacency matrix $C_{\mathcal{G}}$ of the undirected graph $\mathcal{G}$, as well as the probability mass functions of the source symbols $\hat{f}_{n}(\hat{x}), \; \forall n$, and the probability mass functions of the correlation noise variables $g_{m}(w), \; \forall m$. The algorithm starts with the initialization of the messages. All the messages from the check nodes are initialized to one. The messages $q_{nl}(a)$, $a\in\mathbb{F}_{q}$, from the variable nodes are initialized with the prior probability mass functions of the source symbols
 \begin{equation}
 	q_{nl}(a) = \hat{f}_n(a)
	\label{eq:B3_ch3}
 \end{equation}
If the prior distributions are unknown, they are set to be uniform over $\hat{\mathcal{X}}$.

Once the initialization step is completed, the algorithm proceeds with iterative message passing rounds among the factor graph nodes. In particular, at every iteration of the algorithm, beliefs are formed in the nodes and the variable node messages $q_{nl}(a)$, ${a\in\mathbb{F}_{q}}$, are updated as follows:
  \begin{equation}
 	q_{nl}(a) = \alpha_{nl}\prod_{l^\prime\in \mathcal{L}(n)\backslash l}r_{l^\prime n}(a)
	\label{eq:B4_ch3}
 \end{equation}
To elaborate, the message $q_{nl}(a)$ sent from the variable node $n$ to the check node $l$ is updated by multiplying the messages $r_{l^\prime n}(a)$ previously received from all the check nodes $l^\prime\in \mathcal{L}(n)\backslash l$ that are neighbors of the variable node $n$ except for the node $l$. The normalization constant $\alpha_{nl}$ is computed by setting $\sum_{a=0}^{q-1}q_{nl}(a)=1$, so that $q_{nl}(a)$, ${a\in\mathbb{F}_{q}}$, is a valid probability distribution over $\mathbb{F}_{q}$. The message $q_{nl}(a)$ represents the $n$-th variable node's belief that the value of the $n$-th source is $a$, given the information from all the other check nodes except for the check node $l$.

Once the messages from the variable nodes to the check nodes have been computed, we update the messages in the opposite direction. The messages $r_{ln}(a)$, $a\in\mathbb{F}_{q}$, from the check nodes to the variable nodes are updated as:
 \begin{equation}
 	r_{ln}(a) = \sum_{\{\bm{\hat{x}}: \hat{x}_n = a, \hat{y}_l^\prime = \bm{A}^\prime_{l}\bm{\hat{x}}\}}\prod_{n^\prime\in\mathcal{N}(l)\backslash n}\mu_{n^\prime l}(\hat{x}_{n^\prime})
 	\label{eq:B5_ch3}
 \end{equation}
 where $\bm{A}_l^\prime$ denotes the $l$-th row of the matrix $\bm{A}^\prime$ and
  \begin{equation}
 	\mu_{n^\prime l}(\hat{x}_{n^\prime}) = \begin{cases} g_{n^\prime n}(\hat{x}_{n^\prime}-a)q_{n^\prime l}(\hat{x}_{n^\prime}), & \mbox{if } \rho_{nn^{\prime}} \neq 0  \\
	q_{n^\prime l}(\hat{x}_{n^\prime}), & \mbox{if }   \rho_{nn^{\prime}} = 0
	\end{cases}
	\label{eq:B6_ch3}
 \end{equation}
The message $r_{ln}(a)$ sent from the check node $l$ to the variable node $n$ represents the belief that the value of the $n$-th source symbols is $a$ given (i) the beliefs $q_{n^\prime l}(\hat{x}_{n^\prime})$ of all the neighboring variable nodes $n^\prime\in\mathcal{N}(l)\backslash n$ of the check node $l$ except for the variable node $n$, (ii) the $l$-th  network coding constraint and (iii) the pairwise correlation constraints between the variable node $n$ and all the other neighbors $n^\prime\in\mathcal{N}(l)\backslash n$ of the $l$-th check node. The summation in Eq.~\eqref{eq:B5_ch3} is performed over all the configurations of vectors $\bm{\hat{x}}$ that satisfy the condition on the $l$-th check node and have the value $a$ at the $n$-th position. The product term in Eq.~\eqref{eq:B5_ch3} represents the belief of the $l$-th check node on a specific configuration of variables with the value of the $n$-th variable set to $a$. These beliefs incorporate the prior knowledge on the pairwise correlation between the $n$-th variable and the variables $n^\prime\in\mathcal{N}(l)\backslash n$ through the weights that multiply the messages from the variable nodes $n^\prime\in\mathcal{N}(l)\backslash n$, as it can be seen from Eq.~\eqref{eq:B6_ch3}.  Therefore, higher belief values are assigned to configurations that agree with the pairwise correlation model, while configurations that deviate from the correlation model are assigned lower belief values.

 At the end of each message passing round, the variable nodes form their beliefs on the values of the variables $\hat{X}_n$ and the values of the variables are tentatively set to 
 \begin{equation}
 	\hat{x}_n^* = \underset{a \in \mathbb{F}_{q}}{\operatorname{arg\,max}}\,\prod_{l\in\mathcal{L}(n)} r_{ln}(a)
 	\label{eq:B7_ch3}
 \end{equation}
If these values satisfy the network coding constraints imposed on the reconstruction of the sources, namely $\hat{\bm{y}} = \bm{A}\hat{\bm{x}}^*$ (or, equivalently, $\hat{\bm{y}}^\prime = \bm{A}^\prime\hat{\bm{x}}^*$), the solution is considered to be valid. In this case, the decoding stops and the decoder outputs the corresponding solution. If a valid solution is not found after the maximum number of iterations in the decoder has been reached, the decoder declares an error and sets $\hat{x}_n^*$ to be equal to the expected value of $\hat{X}_n$, {\em i.e.}, $\hat{x}_n^* = \mbox{E}[\hat{X}_n]$.

Algorithm \ref{algo:messagepassing} summarizes the steps of the proposed iterative decoding scheme. We employ the parallel schedule for the message update procedure, which implies that all the messages at the variable nodes are updated concurrently given the messages received from the check nodes at a previous stage. Similarly, all check nodes update and send their messages simultaneously. Note that other message passing schedules can be considered ({\em i.e.,} serial) \cite{SharonSerialMP}; however, they are not studied in this work.

 \begin{algorithm}[t]
 \baselineskip=15pt
	\caption{Proposed message passing decoding algorithm} 
	\label{algo:messagepassing}
	\begin{algorithmic}[1] 

	\STATE \textbf{Input:} \\
		 Matrix $\bm{A}^{\prime}$, 
		 vector $\hat{\bm{y}}^{\prime}$,
		 adjacency matrix $C_{\mathcal{G}}$ of the undirected graph $\mathcal{G}$,
		 pmf of the correlation noise $g_{m}(w),\; \forall m$, and
		 pmf of the source symbols $\hat{f}_{n}(\hat{x}),\;\forall n$.
		 
	\STATE \textbf{Initialization:} \\
	 Initialize sets $\mathcal{L}(n) \gets \{l: a_{ln}^\prime \neq 0\}$ and $\mathcal{N}(l) \gets \{n: a_{ln}^\prime \neq 0\}$\\
	 Initialize messages $q_{nl}(a)\gets \hat{f}_n(a), \forall a\in\mathbb{F}_{q}$ and $r_{ln}(a) \gets 1, \forall a\in\mathbb{F}_{q}$\\
	 Set $k \gets 0$  and define $k_{max}$

	\WHILE{$k < k_{max}$}
		
		\STATE  Update the messages from variable to check nodes \\
		$q_{nl}(a) \leftarrow \alpha_{nl}\prod\limits_{l^\prime\in \mathcal{L}(n)\backslash l}r_{l^\prime n}(a),\; {\forall a\in\mathbb{F}_{q}, \; \forall n, \forall l \in \mathcal{L}(n)}$
		
		\STATE Update the messages from check to variable nodes\\
		$r_{ln}(a) \leftarrow \sum\limits_{\{\bm{\hat{x}}: \hat{x}_n = a, \hat{y}_l^\prime = \bm{A}^\prime_{l}\bm{\hat{x}}\}}\prod\limits_{n^\prime\in\mathcal{N}(l)\backslash n}\mu_{n^\prime l}(\hat{x}_{n^\prime}),\;  \forall a\in\mathbb{F}_{q}, \; \forall l, \forall n \in \mathcal{N}(l)$
		
		\STATE Estimate the values of the source symbols
		 \\ $\hat{x}_n^* \gets \underset{ a\in\mathbb{F}_{q}}{\operatorname{arg\,max}}\,\prod\limits_{l\in\mathcal{L}(n)} r_{ln}(a), \; \forall n$
		
		\IF {$\bm{A}^{\prime} \hat{\bm{x}}^{*} == \hat{\bm{y}}^{\prime}$}
		
			\RETURN $\hat{\bm{x}}^{*}$
			
		\ELSE
		
		 	\STATE $k \gets k+1$
							
		\ENDIF

	\ENDWHILE
	
	\IF {$k == k_{max}$}
		
		\RETURN $\hat{\bm{x}}^{*} = (\mbox{E}[\hat{X}_{1}],\mbox{E}[\hat{X}_{2}],\dots, \mbox{E}[\hat{X}_{N}])^{T}$
	
	\ENDIF
	
	\STATE  \textbf{Output:} Estimation of the sequence of source symbols $\hat{\bm{x}}^{*}$
	
\end{algorithmic}
\end{algorithm}


\subsection{Complexity}
\label{sec:complexity}

We now briefly discuss the computational complexity of the proposed message passing algorithm. The preprocessing of the coding matrix $\bm A$ consists of a Gaussian elimination and a variant of Gaussian elimination both with complexity dominated by $\mathcal{O}(N^3)$. This preprocessing is performed only once for a coding matrix $\bm A$. The outgoing message $\bm{q}_{nl} = (q_{nl}(0), q_{nl}(1),\dots, q_{nl}(q-1))$ of a degree-$d_{v}$ variable node consists of $q$ values that are computed by element-wise multiplication of $d_{v}-1$ incoming messages $\bm{r}_{l^{\prime}n} = ({r}_{l^{\prime}n}(0),{r}_{l^{\prime}n}(1),\dots,{r}_{l^{\prime}n}(q-1))$, $l^{\prime} \in \mathcal{L}(n) \backslash l$. This update step requires $q(d_{v}-2)$ multiplications and it is followed by a normalization step, which requires $q-1$ summations and $q$ divisions. Thus, the total number of operations performed in the variable node, per iteration, is $d_{v}(q(d_{v}-2) + q - 1 + q) $, since the node sends messages on each of the $d_{v}$ outgoing edges. Hence, the computational complexity at every variable node per iteration is dominated by $\mathcal{O}(d_{v}^{2}q)$. In order to calculate the outgoing messages of a degree-$d_{c}$ check node, we first need to update the $d_{c}-1$ vectors $\bm{\mu}_{n^{\prime}l} = (\mu_{n^{\prime}l} (0),\mu_{n^{\prime}l} (1),\dots,\mu_{n^{\prime}l} (q-1))$, which requires $q(d_{c}-1)$ multiplications at most. The sum-product expression in Eq.~\eqref{eq:B5_ch3} is essentially a convolution in a Galois field \cite{FFTfinite} and can be computed using dynamic programming approaches \cite{LDPCqMacKay}. It can be performed with $(d_{c}-1)q^{2}$ multiplications and $(d_{c}-1)q(q-1)$ summations. Thus, the check sum processing requires $2(d_{c}-1)q^{2}$ operations per outgoing edge, while the total number of operations during the check node message update is $2d_{c}(d_{c}-1)q^{2}$ and is dominated by $\mathcal{O}(2d_{c}^{2}q^{2})$. For Galois fields of size $q  = 2^{p}$, the computational complexity of the check node messages update can be further reduced by computing the convolution in the transform domain instead of the probability domain. This can be done by using the 2-Hadamard transform of order $q$ \cite{FFTfinite}, which can be computed with complexity $\mathcal{O}(q\log_{2}q)$. Note that compared to similar existing algorithms \cite{RajawatTWCOMM2012,BassiEUSIPCO12,BarrosISWCS09,CruzISIT11} whose complexity with respect to the Galois field size is $\mathcal{O}(q^{d_c})$, $d_c >> 2$, our algorithm achieves a significantly reduced complexity of check node processing, which is only $\mathcal{O}(q^2)$. Overall, the computational complexity of the decoding algorithm is $\mathcal{O}(k(Nd_{v}^{2}q + L^\prime 2d_{c}^{2}q^{2})+2N^3)$, where $N$ is the number of variable nodes, $L^\prime$ is the number of check nodes (equal to the number of rows in the coding matrix $\bm A^\prime$) and $k$
 is the number iterations.


\section{Belief propagation decoding results}
\label{sec:results}


\subsection{Decoding performance with synthetic signals}
\label{sec:synthetic}

We first evaluate the performance of our iterative decoding algorithm with synthetic signals. We study the impact of various system parameters on the decoding performance. For this purpose, we adopt the scenario introduced in \cite{BarrosFactorTrees}. We assume that $N$ sensors are distributed over a geographical area and measure some physical process. The sensors generate continuous real-valued signal samples $s_{n}$, $n= 1,2,\dots,N$, which are quantized with a uniform quantizer $Q[\cdot]$. The quantized samples $x_{n} = Q[s_{n}]$ are then transmitted using network coding to a receiver following the framework given in Fig.~\ref{fig:framework}. 

The vector $\bm{s} = (s_{1}, s_{2}, \dots s_{N})^{T}$ of source values is assumed to be a realization of a $N$-dimensional random vector with multivariate normal distribution $\bm{\mathcal{N}}(\bm{0},\bm{\Sigma})$. The covariance matrix $\bm{\Sigma}$ is defined as 
\begin{equation}
 	\bm{\Sigma} = \left[
	\begin{array}{ c c c c c }
	1 & \rho_{12} & \rho_{13} & \dots & \rho_{1N} \\
	\rho_{21} & 1 & \rho_{23} & \dots & \rho_{2N} \\
	\vdots & \vdots & \vdots & \ddots & \vdots \\ 
	\rho_{N1} & \rho_{N2} & \rho_{N3} & \dots & 1
	\end{array} \right]
\label{eq:C1_ch3}
\end{equation}
The values of the correlation coefficients $\rho_{ij}$ decay exponentially with the distance $d_{ij}$ between the sensors and are calculated using the formula $\rho_{ij} = e^{-\beta d_{ij}}$, where $\beta > 0$. The constant $\beta$ controls the amount of correlation between sensor measurements for a given realization of the sensor network. The undirected graph $\mathcal{G}$ that represents the pairwise correlation between the sources is fully connected since the correlation coefficient is non-zero for all pairs of sources. The probability mass function $g_{m}(w)$ of the correlation noise $W_{m}$,  for a pair of sources $S_{i}$ and $S_{j}$ is computed analytically by integrating the joint probability distribution $p(s_{i},s_{j})$ over the quantization intervals in order to obtain the joint pmf of the quantized samples, and then by summing over all configurations that  yield the same correlation noise
\begin{equation}
\begin{split}
	g_{m}(w) &= \sum\limits_{x_{i},x_{j}:x_{i}-x_{j} = w}f(x_{i},x_{j}) 
	=\sum\limits_{x_{i},x_{j}:x_{i}-x_{j} = w}\int\limits_{I(x_{i})} \int \limits_{I(x_{j})} p(s_{i},s_{j})ds_{i}ds_{j}
\end{split}
\label{eq:C2_ch3}
\end{equation}
The marginal probability mass functions $f_{n}(x)$ are obtained in a similar way. 

We first study the influence of the correlation on the decoding performance. We distribute $N=20$ sensors uniformly over a unit square and vary the value of $\beta$ that controls directly the correlation among the sensor measurements. We choose $\beta = 0.01$, $0.05$, $0.1$ and $0.2$. For these values of $\beta$, the correlation coefficients of all possible pairs of sources belong to the intervals $[0.9898,\; 0.9997]$, $[0.9502, \; 0,9986 ]$, $[0.9027, \; 0.9972]$ and $[0.8153,\; 0.9943]$, respectively. The sensor measurements are quantized using a 3-bit and respectively a 4-bit uniform quantizer. The maximum number of iterations in the decoding algorithm is set to $k_{max} = 100$ and the number of transmitted source sequences is set to $N_{samp} = 20000$, unless stated otherwise.

\begin{figure*}[t]
	\begin{center}
			\subfloat[$q = 8$]{\label{fig:ErrorVsCorrelationM3}\includegraphics[width=0.46\textwidth]{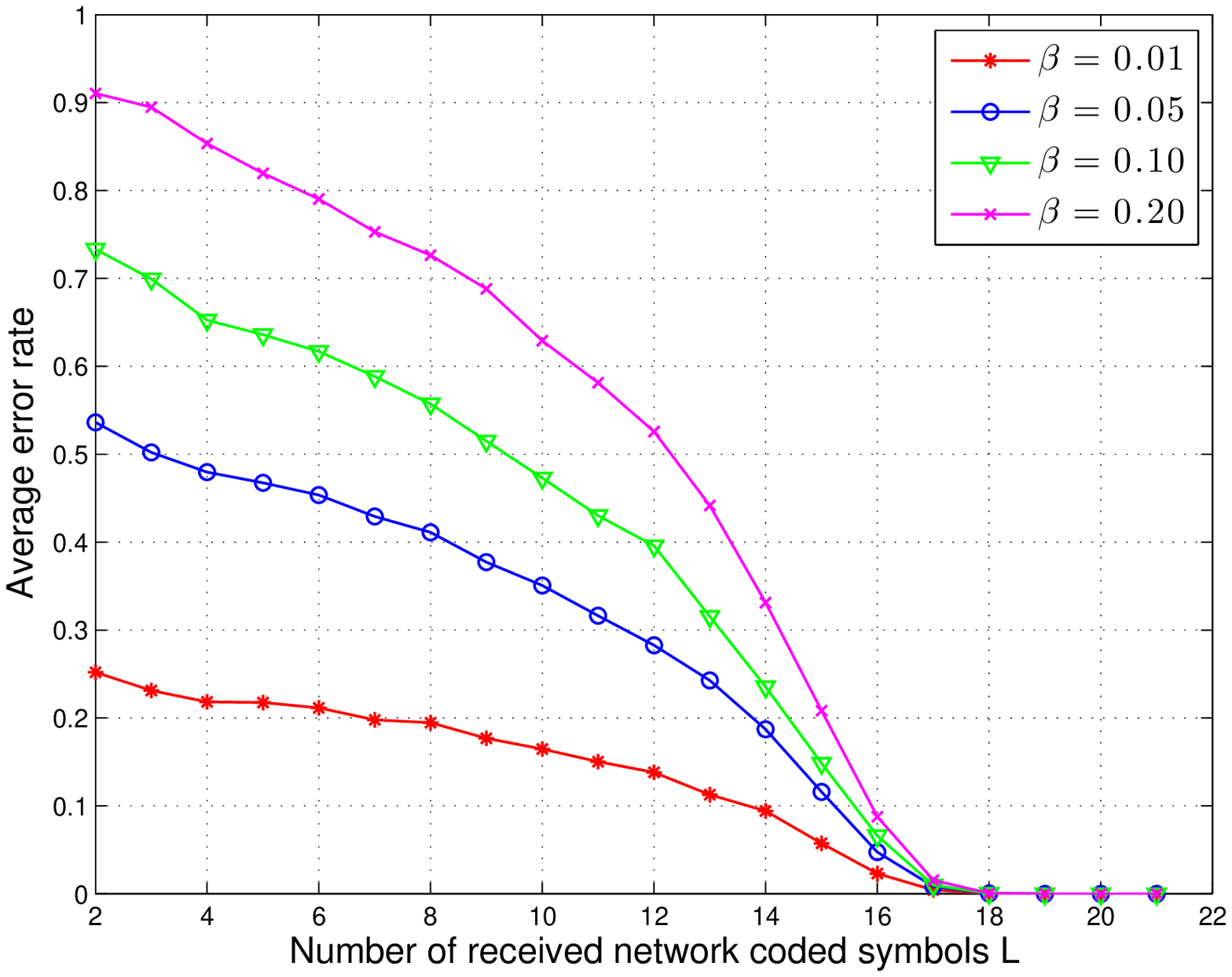}} \quad \quad 
			\subfloat[$q = 16$]{\label{fig:ErrorVsCorrelationM4}\includegraphics[width=0.46\textwidth]{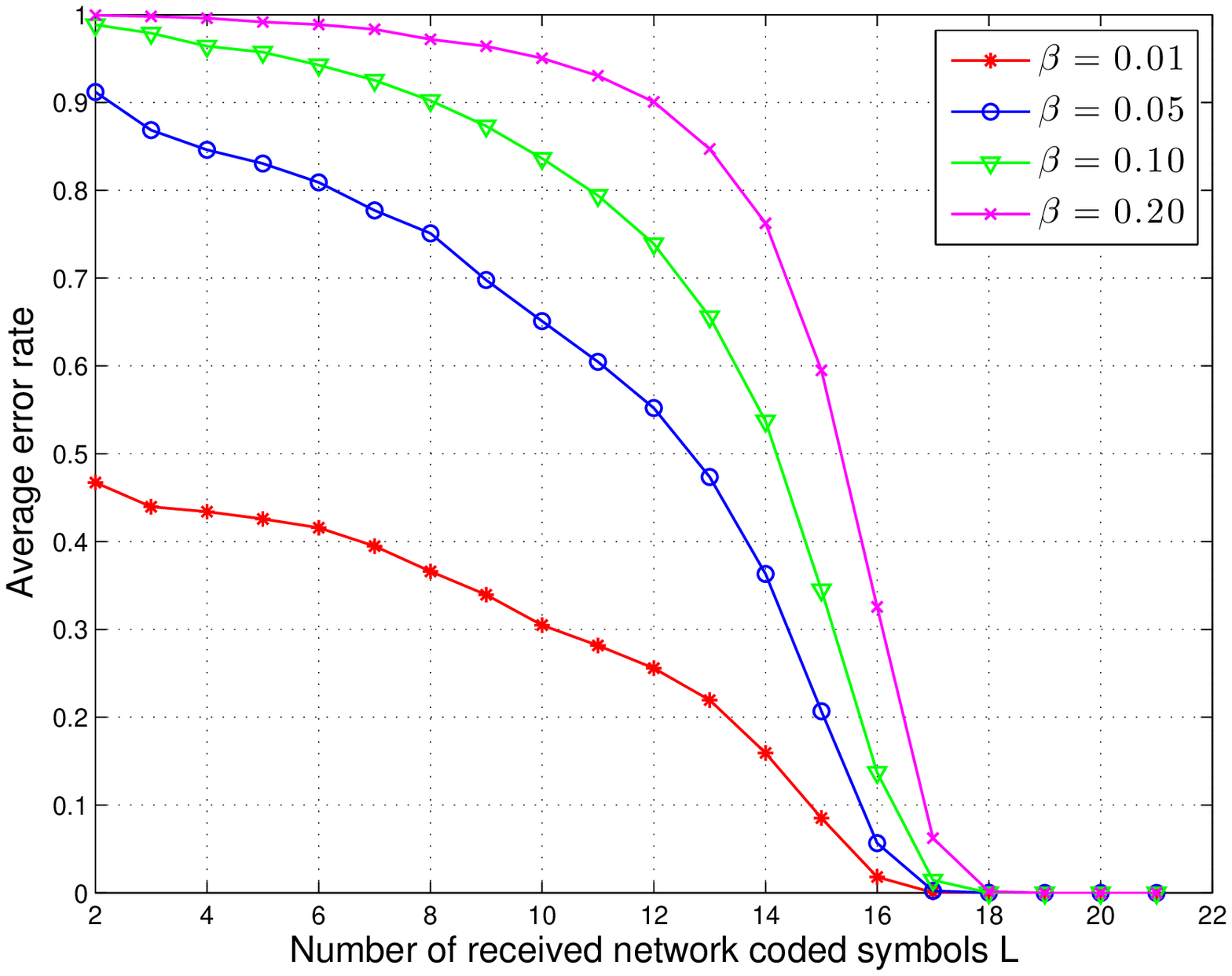}}
	\end{center}
	\caption{Average decoding error rate versus the number of received network coded symbols $L$ for a sensor network with $N = 20$ sensors and various values of the parameter $\beta$. The sensor measurements are sampled from a multivariate normal distribution and are quantized with (a) a 3-bit ($q = 8$) uniform quantizer and (b) a 4-bit ($q = 16$) uniform quantizer.  \label{fig:syntheticdataErrorvsBeta}} 
\end{figure*}

Fig.~\ref{fig:syntheticdataErrorvsBeta} shows the decoding performance in terms of average error rate with respect to the number of received network coded symbols for finite field sizes $q = 8$ and $q = 16$. In this first set of experiments, the cardinality of the source alphabet, as well as the size of the finite field used for data representation and network coding operations, is equal to the number of quantization bins of the uniform quantizer. The average error rate is defined as the ratio between the number of erroneously decoded source sequences and the number of transmitted source sequences $N_{samp}$. We can see that the decoder may tolerate a few missing network coded symbols and still provide perfect reconstruction of the transmitted source sequences. In these cases, the correlation between the sources compensates for the missing network coded data. As the number of network coded symbols available at the decoder reduces, the performance of the decoder gradually deteriorates. The importance of this performance degradation is dependent on the correlation among the source symbols. We can observe that the stronger the correlation between the sources ({\em i.e.}, the smaller the values of the parameter $\beta$), the less network coded symbols are required at the decoder to achieve a given decoding error rate. We also remark that the quantization has an impact on the correlation between the transmitted source symbols. For source data sampled from the same joint probability distribution  $\bm{\mathcal{N}}(\bm{0},\bm{\Sigma})$, finer quantization leads to larger values of the correlation noise, which results in worse decoding performance. This can be observed in Fig.~\ref{fig:syntheticdataErrorvsBeta} by comparing the decoding performance for the same value of the parameter $\beta$ and for different values of Galois field size $q$. In particular, for a given value of $\beta$ and for the same number of received network coded symbols $L$, the average error rate is higher when the source data is represented in larger Galois fields ({\em i.e.}, when finer quantization is applied). 

We investigate next the influence of the finite field size $q$ on the decoding performance when the field size is not necessarily equivalent to the cardinality of the quantized sources. We use the same sensor network topology as previously with $N = 20$ sensors distributed uniformly over a unit square. The sensor measurements are quantized using a 3-bit uniform quantizer. Thus, the cardinality of the source alphabet is $q^{\prime} = 8$. We vary the size $q$ of the Galois field that is used for representing the source data and performing the network coding operations in the network. The maximum number of iterations at the decoder is set to $k_{max}  = 100$. The results are obtained by transmitting $N_{samp} = 10000$ source sequences.

\begin{figure*}[t]
	\begin{center}
			\subfloat[$\beta = 0.01$]{\label{fig:ErrorVsFieldSizeBeta1}\includegraphics[width=0.46\textwidth]{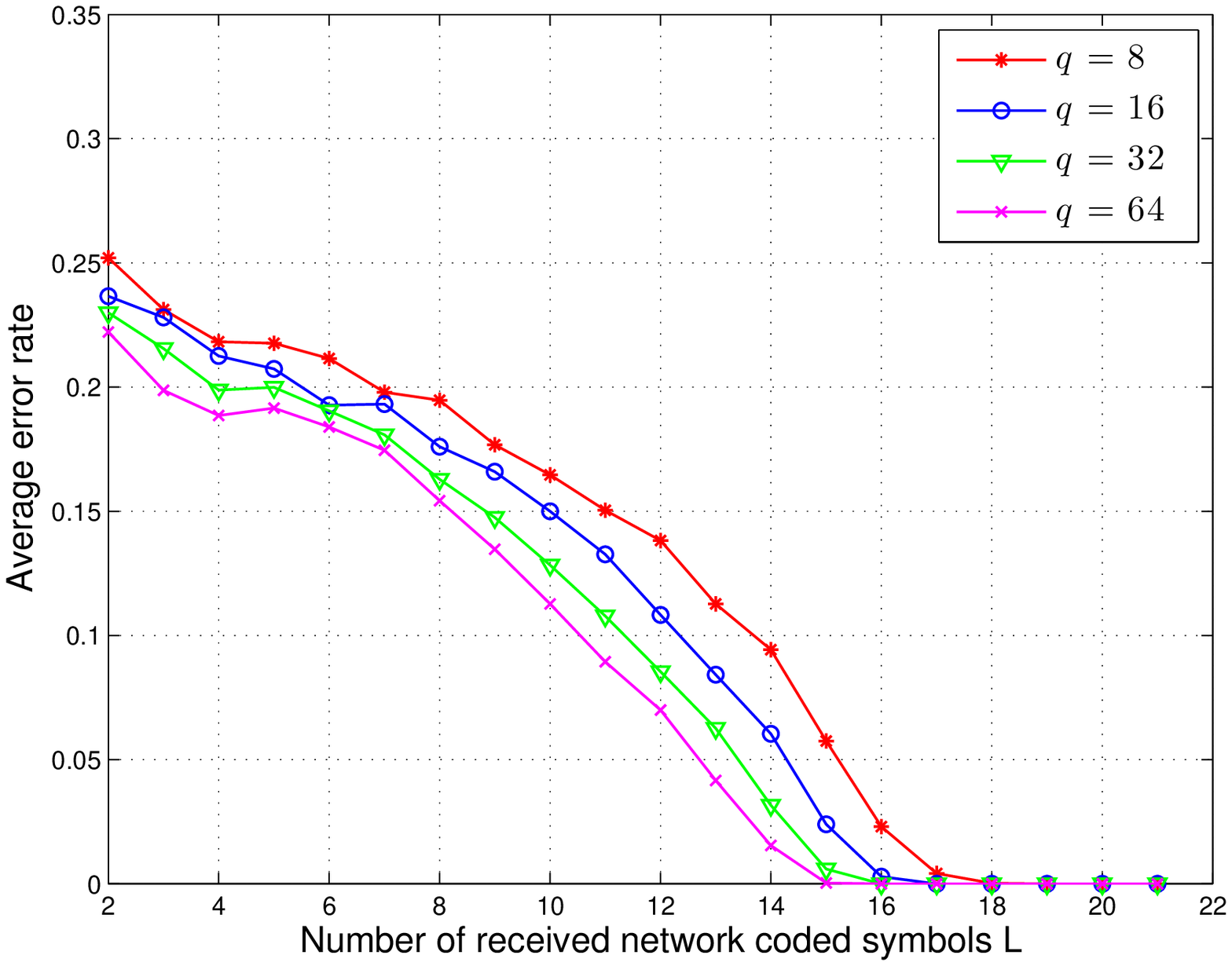}}  \quad \quad 
			\subfloat[$\beta = 0.05$]{\label{fig:ErrorVsFieldSizeBeta2}\includegraphics[width=0.46\textwidth]{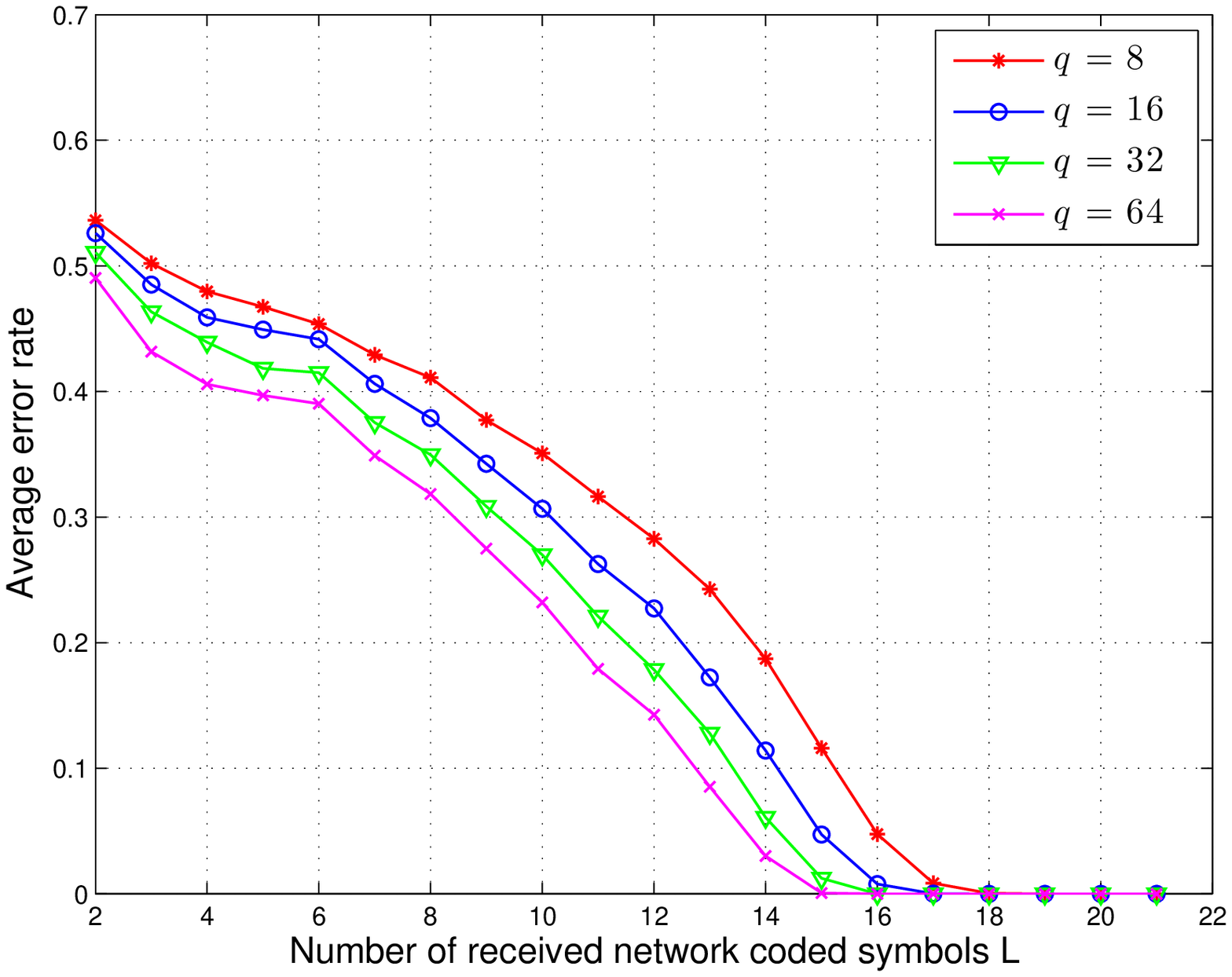}}
	\end{center}
	\caption{ Average decoding error rate versus the number of received network coded symbols $L$ for a sensor network with $N = 20$ sensors and for various values of the Galois field size $q$, used for data representation and network coding operations. The sensor measurements are sampled from two different multivariate normal distributions ($\beta = 0.01$ and $\beta = 0.05$) and are quantized with a 3-bit ($q^{\prime} = 8$) uniform quantizer. \label{fig:ErrorVsFieldSize}}
\end{figure*}

In Fig.~\ref{fig:ErrorVsFieldSize}, we present the average error rate versus the number of received network coded symbols $L$ for different values of Galois field size and two values of the parameter $\beta$, which controls the correlation among the sources. We can see that the decoding performance for a given number of received symbols $L$ improves when larger Galois fields are used for the source data representation. The employment of larger Galois fields for the data representation introduces more diversity in the coding operations, which results in numerous candidate solutions in the decoder that are not valid according to the source data model. Thus, the decoder is able to eliminate these solutions during the decoding process and the decoding performance is improved. Recall that the same behavior has been observed in the upper bound on the decoding error probability under MAP decoding rule (see Fig.~\ref{fig:bound}(b)). It is worth noting that this performance improvement comes at the cost of a larger number of bits that have to be transmitted in the network.


\subsection{Decoding of correlated images}
\label{sec:images}

We now illustrate the behavior of our iterative decoding algorithm in the problem of decoding correlated images from an incomplete set of network coded packets. For our tests we use the video sequences {\em Silent} and {\em Foreman} in QCIF format. Each image in the sequences corresponds to a $144\times 176$ grayscale image with pixel values in the range $[0,255]$. We extract the first $N =15$ consecutive images from each sequence and assume that the data transmitted by each source in Fig.~\ref{fig:framework} corresponds to one of these images. In order to obtain source data with different alphabet sizes, we quantize the images by discarding the least significant bits of the binary representation of pixel values. The pixel values generated by the sources are directly mapped to Galois field values and transmitted with network coding to the receiver. At the receiver, the network coded symbols are decoded with the proposed iterative message passing algorithm and mapped back to pixel values. The maximum number of iterations at the decoder is set to $k_{max} = 100$.

The temporal correlation between the images in the video sequences is modeled with a zero mean discrete Laplacian distribution \cite{discreteLaplace}. This model is widely used in Distributed Video Coding (DVC) \cite{noisemodelingDVC}, where the correlation noise between the original frame and the side information is considered to have a Laplacian distribution. The probability mass function of the correlation noise $W_{m}$ between two sources (images) $S_{i}$ and $S_{j}$ is given by
\begin{equation}
	g_{m}(w) = \frac{1-p_{m}}{1+p_{m}}p_{m}^{|w|}, \quad p_{m} \in (0,1)
	\label{eq:C3_ch3}
\end{equation}
Only one parameter per pair of correlated sources is required in order to model the correlation noise. The magnitude of the parameter $p_{m}$ determines the degree of correlation between the sources $S_{i}$ and $S_{j}$. For small values of the parameter $p_{m}$, the probability distribution is concentrated around zero, which means that the sources $S_{i}$ and $S_{j}$ are highly correlated. On the contrary, a large parameter $p_{m}$ indicates that the corresponding sources are not strongly correlated.

\begin{figure}[t]
	\begin{center}
			\subfloat[Average error rate]{\label{fig:ImgErrorRate}\includegraphics[width=0.46\textwidth]{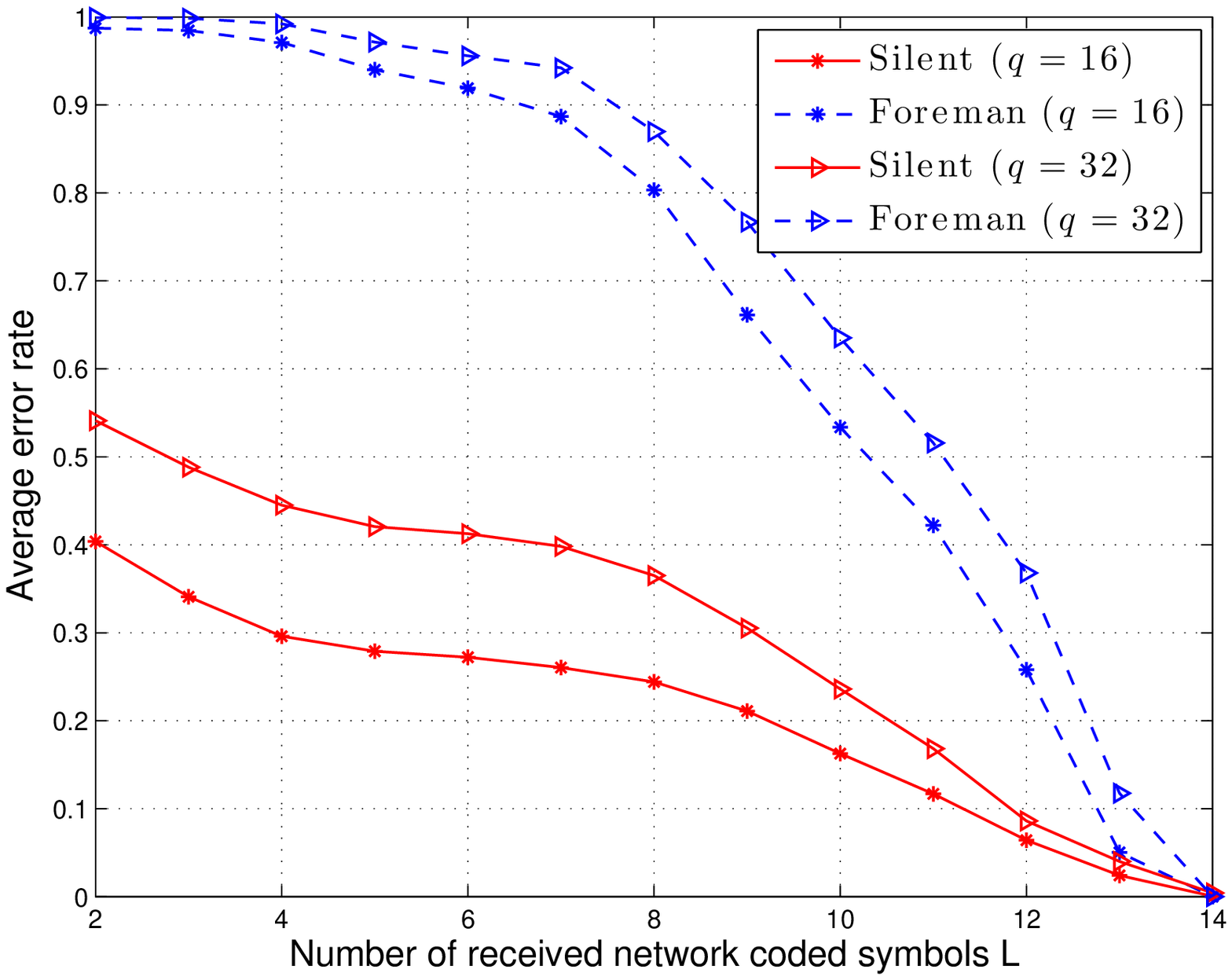}} \quad \quad 
			\subfloat[Average PSNR]{\label{fig:ImgPSNR}\includegraphics[width=0.46\textwidth]{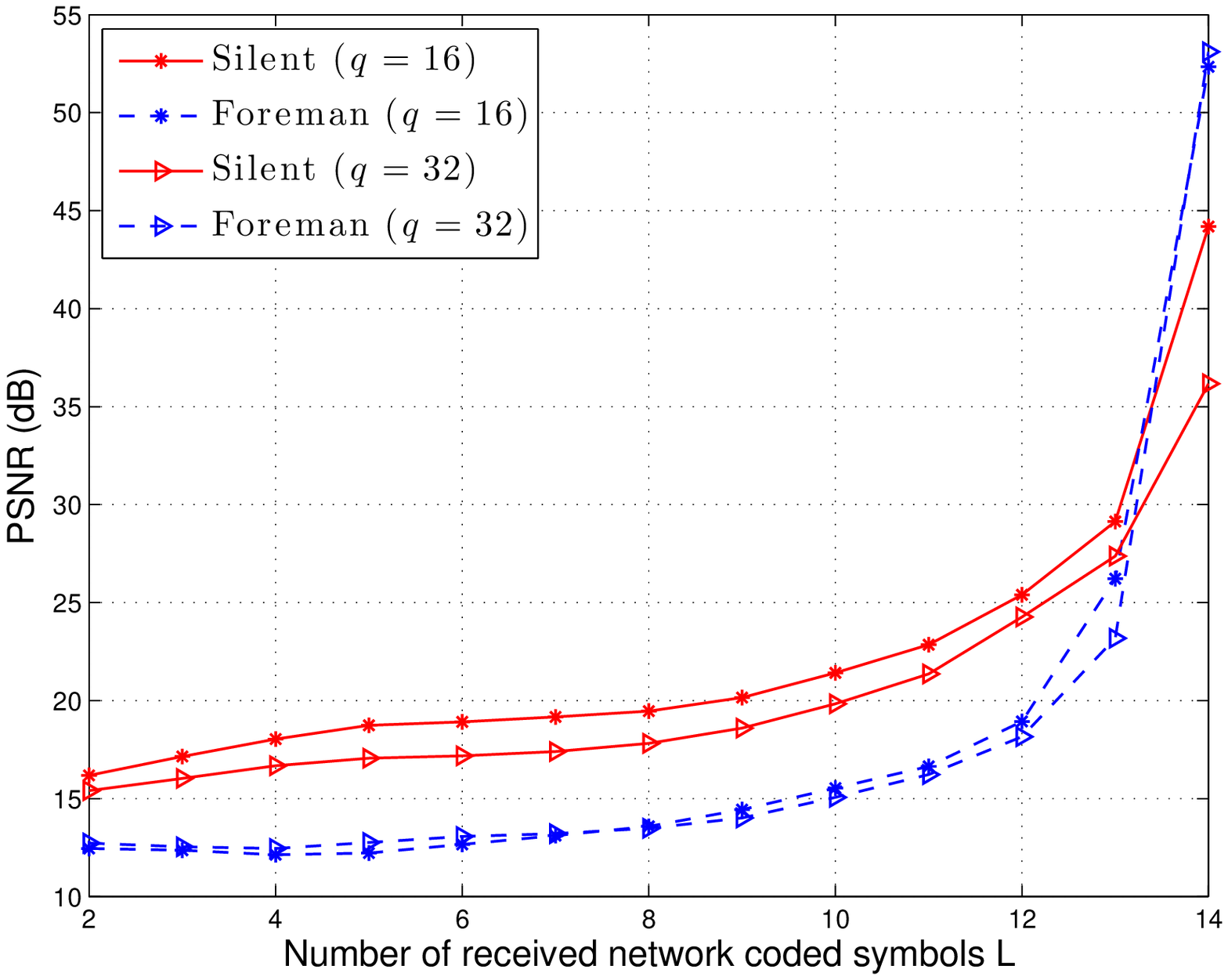}}
	\end{center} 
	\caption{ Decoding performance in terms of (a) average error rate and (b) average PSNR as a function of the number of received network coded symbols $L$ for $N = 15$ correlated images extracted from the {\em Silent} and the {\em Foreman} QCIF sequences. The images are quantized to 4 bits and 5 bits prior to transmission and the coding operations are performed in the fields of size $q = 16$ and $q = 32$, respectively. \label{fig:Images}} \vspace{-0.4cm}
\end{figure}

Fig.~\ref{fig:Images} shows the decoding performance in terms of average error rate and the average PSNR versus the number of network coded symbols available at the decoder. The original images are quantized to either $n = 4$ or $n = 5$ bits prior to transmission, and the network coding operations are performed in finite fields of size $q = 16$ and $q = 32$, respectively. The average error rate is defined as the ratio between the number of erroneously decoded source sequences and the total number of transmitted source sequences, which is equal to the number of pixels in one image. The PSNR for each decoded image is calculated as
\begin{equation}
	\mbox{PSNR} = 10\log _{10} \frac{(2^{n} - 1)^2}{\mbox{MSE}}
	\label{eq:C4_ch3}
\end{equation}
where MSE is the mean square error between the original quantized image and the decoded image, and $n$ is the number of bits used to represent the pixel value. We can see that the decoding performance is significantly worse for the {\em Foreman} sequence as compared to the {\em Silent} sequence. This is due to the higher motion present in the {\em Foreman} sequence, which leads to lower correlation between pairs of frames. On the contrary, the {\em Silent} sequence is more static and motion is present only in limited regions of the images. Thus the correlation between pairs of frames is larger; that leads to better performance both in terms of average error rate and average PSNR.

\begin{figure}[t]
	\begin{center}
		\subfloat{\includegraphics[width=0.29\textwidth]{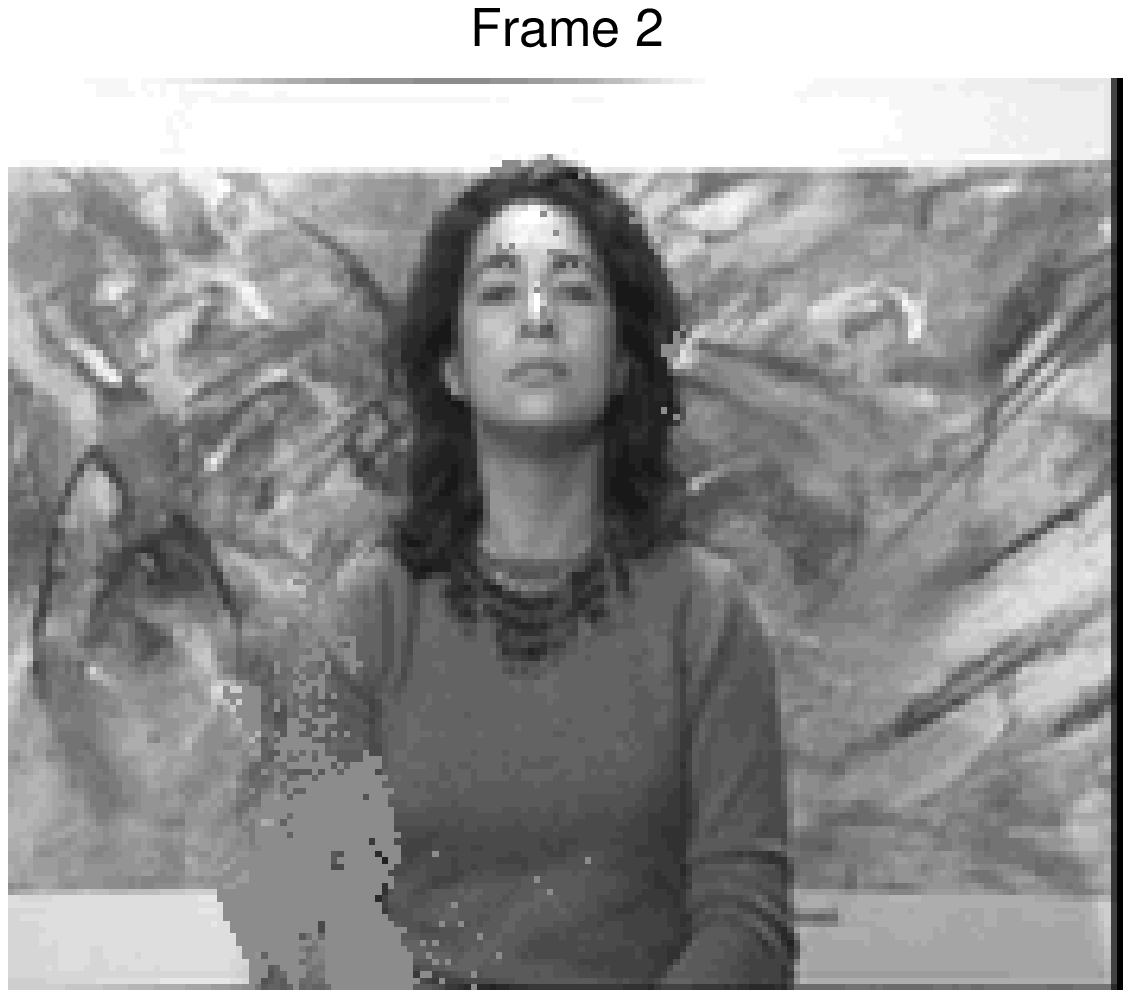}}~~~~~
		\subfloat{\includegraphics[width=0.29\textwidth]{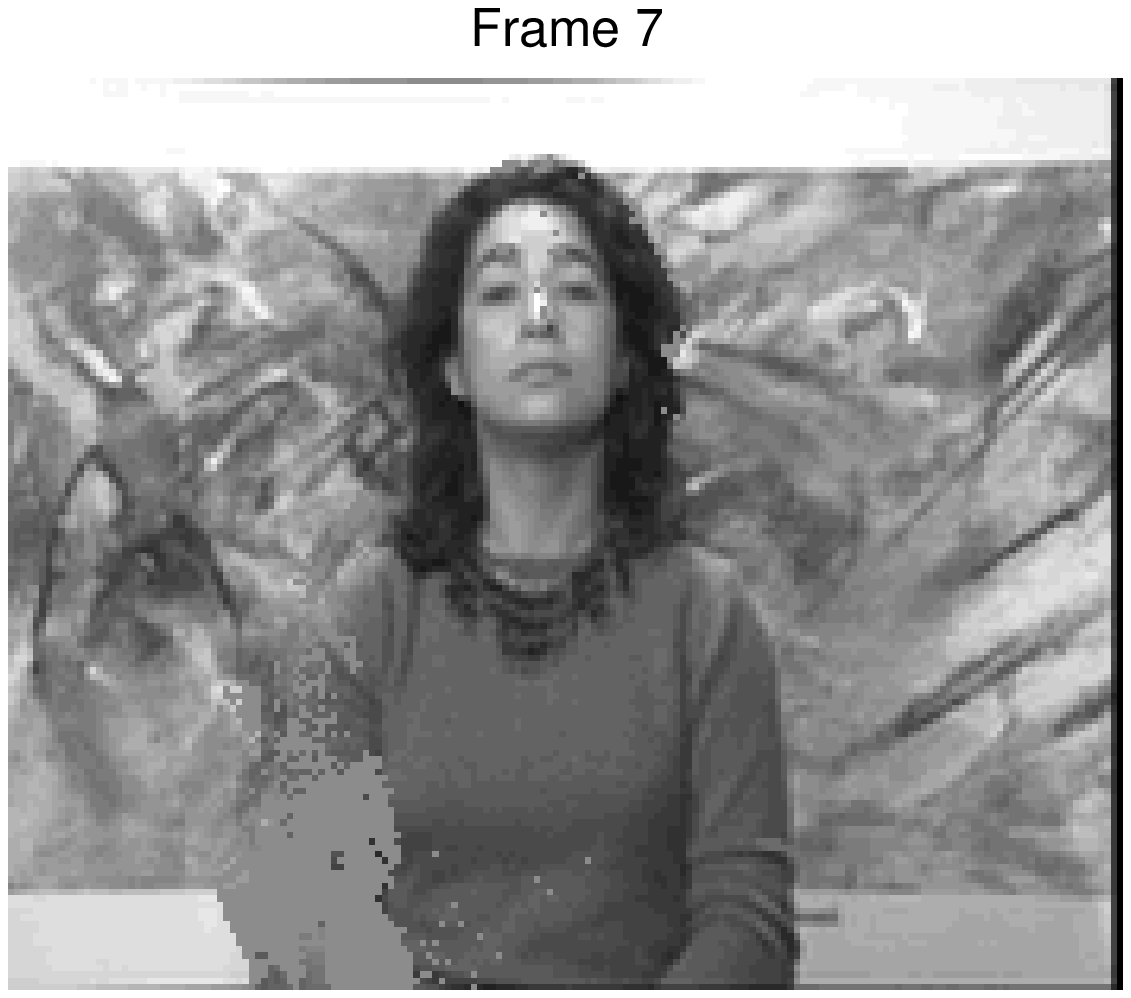}}~~~~~
		\subfloat{\includegraphics[width=0.29\textwidth]{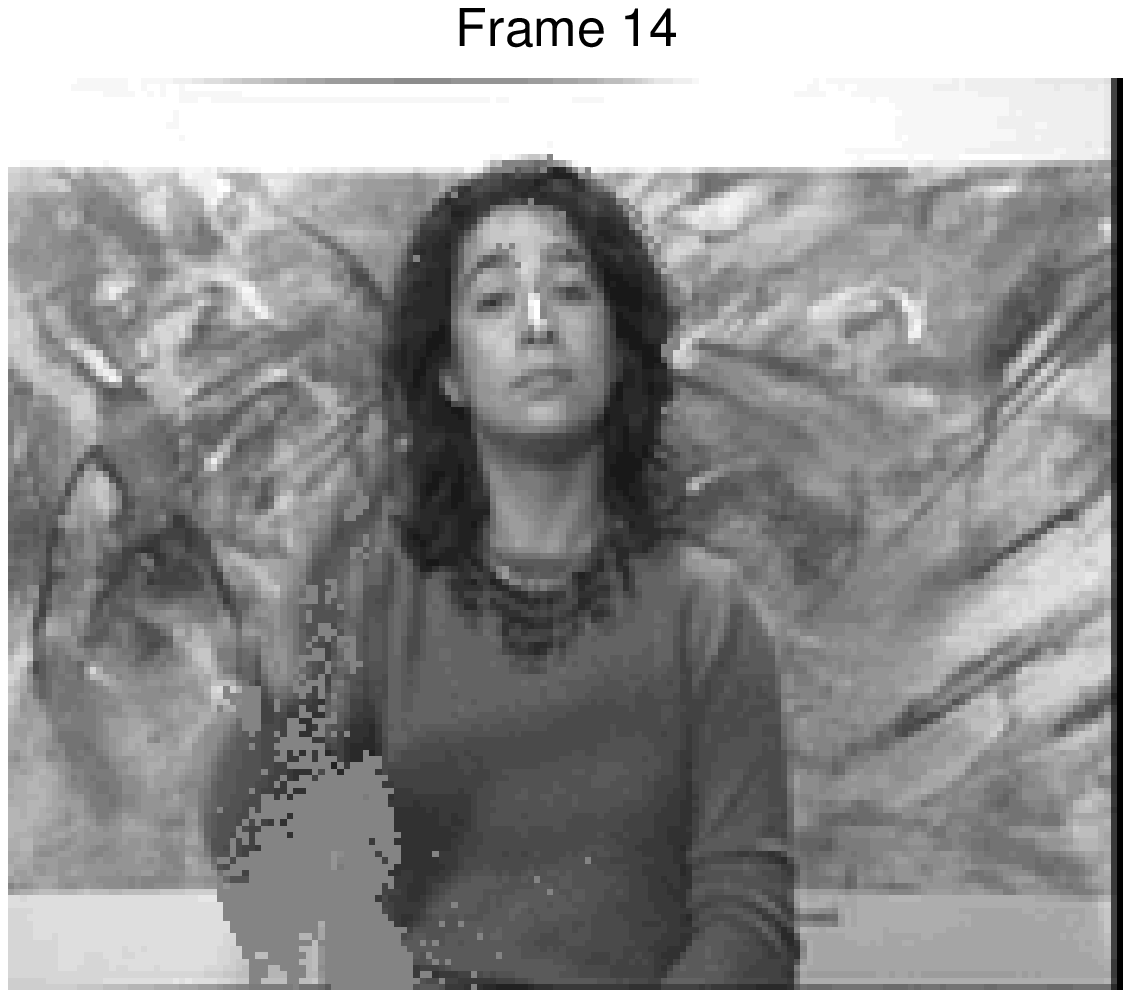}}
	\end{center}
	\caption{Decoded frames of the  {\em Silent} QCIF video sequence. The images are quantized to 5 bits prior to transmission and the coding operations are performed in a field of size $q = 32$. The decoding is performed from $L = 13$ network coded symbols.
	\label{fig:visualperf}}
	\vspace{-0.4cm}
\end{figure}

In Fig.~\ref{fig:visualperf} we illustrate the visual quality of some of the decoded images for the {\em Silent} sequence. The images are quantized to 5 bits ($q = 32$) and the decoding is performed from $L=13$ network coded symbols. We can observe that the majority of the errors appear in the regions that are characterized by significant motion, where the simple correlation model losses its accuracy.


\section{Conclusions}
\label{sec:conclusions}

We have studied the problem of decoding network coded data, when the number of received network coded symbols is not sufficient for perfect reconstruction of the original source data with conventional decoding methods. We have analyzed the performance of a {\em maximum a posteriori} decoder that approximates the source data with the most probable source symbol sequence given the incomplete set of network coded symbols and the matrix of coding coefficients. In particular, we have derived the upper bound on the probability of decoding error under the MAP decoding rule and established the sufficient conditions on the number of network coded symbols required for achieving a decoding error probability below a certain value. The theoretical analysis has shown that the decoding performance improves as the correlation among the source symbols increases. We have also proposed a practical algorithm for decoding network coded data from an incomplete set of data once prior information about data correlation is available at decoder. Our algorithm is based on iterative message passing and jointly considers the network coding and the correlation constraints expressed in finite and real fields, respectively. We have demonstrated the performance of our algorithm through simulations on synthetic signals and on sets of correlated images extracted from video sequences. The results show the great potential of such decoding methods for the recovery of source data when exact reconstruction is not feasible due to the lack of network coded information. It is worth noting that conventional decoding methods such as Gaussian elimination fail to provide any reconstruction when some of the network coding symbols are missing at the decoder. Our method however is able to partially recover the source information even from incomplete network coded data.

\begin{appendices}
\section{Proof of Proposition \ref{prop:1}}
\label{app:1}

\begin{proof}

When a vector $\hat{\bm x}$ of source symbols is transmitted, a vector of network coded symbols $\hat{\bm y} = \bm{A}\hat{\bm x}$ is observed at the decoder, where the entries of  $\bm{A}$ are i.i.d. uniform random variables in $\mathbb{F}_{q}$. An error occurs in the MAP decoder of Eq.~\eqref{eq:D2_ch3} when there exists a sequence $\hat{\bm w}\in \hat{\mathcal{X}}^{N}$, with $\hat{\bm w} \neq \hat{\bm x}$, such that the probability $\hat{f}(\hat{\bm w})\geq \hat{f}(\hat{\bm x})$ and $\bm{A}\hat{\bm w} =\hat{\bm y}$. Therefore, we can write the conditional probability of error as the probability of the union of events where $\bm{A}\hat{\bm w} =\hat{\bm y}$, for all vectors $\hat{\bm w} \neq \hat{\bm x}$ such that $\hat{f}(\hat{\bm w})\geq \hat{f}(\hat{\bm x})$
\begin{equation}
	\mbox{Pr}\{\mbox{error}|\hat{\bm x}\} = \mbox{Pr}\Big\{\bigcup _{\substack{\hat{\bm w }\in \hat{\mathcal{X}}^{N},\hat{\bm w} \neq \hat{\bm x}: \\ \hat{f}(\hat{\bm w})\geq \hat{f}(\hat{\bm x})}}\bm{A}\hat{\bm w} = \hat{\bm y}\Big\}
	\label{eq:D5_ch3}
\end{equation}
 Applying the union bound and using the property that for any set of events $\{B_{i}\}$
\begin{equation}
	\mbox{Pr}\Big\{\bigcup\limits_{i}B_{i}\Big\} \leq \Big{[}\sum\limits_{i}\mbox{Pr}\{ B_{i}\}\Big{]}^{\rho}
	\label{eq:D6_ch3}
\end{equation}
 for any $0\leq \rho \leq 1$ \cite{GallagerITRCbook},  Eq.~\eqref{eq:D5_ch3} yields
\begin{equation}
		\mbox{Pr}\{\mbox{error}|\hat{\bm x}\}  \leq \Big{[}\sum_{\substack{\hat{\bm w }\in \hat{\mathcal{X}}^{N}, 
		\hat{\bm w} \neq \hat{\bm x}: \\ \hat{f}(\hat{\bm w})\geq \hat{f}(\hat{\bm x})}}\mbox{Pr}\{\bm{A}\hat{\bm w} = \hat{\bm y}\}\Big{]}^{\rho}
	\label{eq:D7_ch3}
\end{equation}
The probability that the vector $\hat{\bm{w}} \in \hat{\mathcal{X}}^{N}$ satisfies the network coding constraints, namely the probability that $\bm{A}\hat{\bm w} = \hat{\bm y}$, can be written as
\begin{equation}
\begin{split}
		\mbox{Pr}\{\bm{A}\hat{\bm w} = \hat{\bm y}\} &= \mbox{Pr}\{\bm{A}\hat{\bm w} = \bm{A}\hat{\bm x}\} 
		=\mbox{Pr}\{\bm{A}(\hat{\bm w}+ \hat{\bm x}) = {\bm 0}\} \\
		& = \mbox{Pr}\{\bm{A}\hat{\bm{z}} = \bm{0}\}  
		 = \mbox{Pr}\Big\{\bigcap\limits_{l = 1}^{L}\Big(\sum\limits_{n = 1}^{N}a_{ln}\hat{z}_{n} = 0 \Big)\Big\} 
	\label{eq:D8_ch3}
\end{split}	
\end{equation}
where $\hat{\bm{z}} = \hat{\bm w}+\hat{\bm x}$, with $\hat{\bm{z}} \in \mathbb{F}_{q}^{N}$. The vector $\hat{\bm{z}}$ has at least one non-zero element, since $\hat{\bm w} \neq \hat{\bm x}$. Since the entries of the coding matrix $\bm{A}$ are i.i.d. uniform random variables in $\mathbb{F}_{q}$ with $\mbox{Pr}\{a_{ln} = \alpha\} = q^{-1}$, $\forall \alpha \in \mathbb{F}_{q}$, it holds that $\mbox{Pr}\{a_{ln}\hat{z}_{n} = \alpha\} = q^{-1}$, $\forall \alpha \in \mathbb{F}_{q}$ and $\forall \hat{z}_{n} \in \mathbb{F}_{q}\backslash 0$. Thus, Eq.~\eqref{eq:D8_ch3} can be further written  as
\begin{equation}
\begin{split}
		\mbox{Pr}\{\bm{A}\hat{\bm w} = \hat{\bm y}\} & = \prod\limits_{l=1}^{L}\mbox{Pr}\Big\{ \sum\limits_{n = 1}^{N}a_{ln}\hat{z}_{n} = 0\Big\} 
		 = \prod\limits_{l=1}^{L} q^{-1} = q^{-L}
	\label{eq:D9_ch3}
\end{split}	
\end{equation}
where $L$ is the number of network coded symbols available at the decoder. Substituting the value that we obtain from Eq.~\eqref{eq:D9_ch3} into Eq.~\eqref{eq:D7_ch3}, and using the Gallager's bounding technique \cite{GallagerITRCbook}, the conditional probability of error is upper bounded by
\begin{equation}
\begin{split}
		\mbox{Pr}\{\mbox{error}|\hat{\bm x}\}  &\leq \Big{[}\sum_{\substack{\hat{\bm w }\in \hat{\mathcal{X}}^{N}, 
		\hat{\bm w} \neq \hat{\bm x}: \\ \hat{f}(\hat{\bm w})\geq \hat{f}(\hat{\bm x})}}q^{-L}\Big{]}^{\rho} 
		\overset{(a)}{\leq}  \Big{[}\sum_{\substack{\hat{\bm w }\in \hat{\mathcal{X}}^{N}, 
		\hat{\bm w} \neq \hat{\bm x}: \\ \hat{f}(\hat{\bm w})\geq \hat{f}(\hat{\bm x})}}\Big(\frac{\hat{f}(\hat{\bm w})}{\hat{f}(\hat{\bm x})}\Big{)}^		{\frac{1}{1+\rho}}q^{-L}\Big{]}^{\rho} \\
		 &=  q^{-\rho L}\hat{f}(\hat{\bm x})^{-\frac{\rho}{1+\rho}}\Big{[}\sum_{\substack{\hat{\bm w }\in \hat{\mathcal{X}}^{N}, 
		\hat{\bm w} \neq \hat{\bm x}: \\ \hat{f}(\hat{\bm w})\geq \hat{f}(\hat{\bm x})}}\hat{f}(\hat{\bm w})^{\frac{1}{1+\rho}}\Big{]}^{\rho}
		 \overset{(b)}{\leq} q^{-\rho L}\hat{f}(\hat{\bm x})^{-\frac{\rho}{1+\rho}}\Big{[}\sum \limits_{\hat{\bm w }\in \hat{\mathcal{X}}^{N}}\hat{f}(\hat{\bm w})	^{\frac{1}{1+\rho}}\Big{]}^{\rho}
	\label{eq:D10_ch3}	
\end{split}
\end{equation}

\noindent where inequality (a) results from multiplying each term of the summation by ${\Big{(}\frac{\hat{f}(\hat{\bm w})}{\hat{f}(\hat{\bm x})}\Big{)}^{\frac{1}{1+\rho}} \geq1}$, while the right side of (b) is obtained be relaxing the constraint $\hat{f}(\hat{\bm w})\geq \hat{f}(\hat{\bm x})$ in the summation and by summing over all $\hat{\bm w }\in \hat{\mathcal{X}}^{N}$. 
Finally, the average error probability $P_{e}$ is 

\begin{align}
	P_{e} &= \sum\limits_{\hat{\bm x}\in \hat{\mathcal{X}}^{N}}\mbox{Pr}\{\mbox{error}|\hat{\bm x}\} \hat{f}(\hat{\bm x}) 
	 \leq \sum\limits_{\hat{\bm x}\in \hat{\mathcal{X}}^{N}}q^{-\rho L}\hat{f}(\hat{\bm x})^{-\frac{\rho}{1+\rho}}\Big{[}\sum \limits_{\hat{\bm w }\in \hat{\mathcal{X}}^{N}}\hat{f}(\hat{\bm w})^{\frac{1}{1+\rho}}\Big{]}^{\rho}\hat{f}(\hat{\bm{x}})  \nonumber \\
	&= q^{-\rho L}\sum\limits_{\hat{\bm x}\in \hat{\mathcal{X}}^{N}}\hat{f}(\hat{\bm x})^{\frac{1}{1+\rho}}\Big{[}\sum \limits_{\hat{\bm w }\in \hat{\mathcal{X}}^{N}}\hat{f}(\hat{\bm w})^{\frac{1}{1+\rho}}\Big{]}^{\rho} 
	= q^{-\rho L}\Big{[}\sum\limits_{\hat{\bm x}\in \hat{\mathcal{X}}^{N}}\hat{f}(\hat{\bm x})^{\frac{1}{1+\rho}}\Big{]}\Big{[}\sum \limits_{\hat{\bm w }\in \hat{\mathcal{X}}^{N}}\hat{f}(\hat{\bm w})^{\frac{1}{1+\rho}}\Big{]}^{\rho} \label{eq:D11_ch3} \\
	& = q^{-\rho L} \Big{[}\sum\limits_{\hat{\bm x}\in \hat{\mathcal{X}}^{N}}\hat{f}(\hat{\bm x})^{\frac{1}{1+\rho}}\Big{]}^{1+\rho}
	 = q^{-\rho L} \Big{[}\sum\limits_{{\bm x}\in \mathcal{X}^{N}}f({\bm x})^{\frac{1}{1+\rho}}\Big{]}^{1+\rho}  = 2^{-\rho L \log_{2}q + (1+\rho)\log_{2} \Big{[}\sum\limits_{{\bm x}\in \mathcal{X}^{N}}f({\bm x})^{\frac{1}{1+\rho}}\Big{]}}\nonumber	\\
	& = 2^{-\rho L\log_2 q +\rho H_\rho (\bm X) - D_{KL} (f_\rho(\bm x)|| f(\bm x))} \nonumber
\end{align}
%
Since the inequality in Eq.~\eqref{eq:D11_ch3} holds for any $0\leq\rho\leq 1$, we obtain the bound in Eq.~\eqref{eq:D4_1_ch3}.
\end{proof}

\section{Proof of Proposition \ref{prop:2}}
\label{app:2}
We first prove that the function 
\begin{equation}
	F(r) = r\log_{2}\Big{[}\sum_{{\bm x}\in \mathcal{X}^{N}}f(\bm x)^{\frac{1}{r}}\Big{]}
	\label{eq:D13_ch3}
\end{equation}
\noindent is a convex function of $r$,  $r  >  0$, with strict convexity if $f(\bm{x})$ is not uniform over $\mathcal{X}^{N}$. To show that, it is sufficient to show that for any $s,t > 0$ and $0 < \lambda < 1$
\begin{equation}
	F( \lambda s + (1-\lambda )t) \leq \lambda F(s) + (1 - \lambda) F(t)
	\label{eq:D14_ch3}
\end{equation}
Let us set $r = \lambda s + (1-\lambda )t $. We have 
\begin{equation}
\begin{split}
	\sum_{{\bm x}\in \mathcal{X}^{N}}f(\bm x)^{\frac{1}{r}} =  \sum_{{\bm x}\in \mathcal{X}^{N}}  f(\bm x)& ^{\frac{\lambda + (1-\lambda)}{r}} 
	 =  \sum_{{\bm x}\in \mathcal{X}^{N}}f(\bm x)^{\frac{\lambda}{r}}  f(\bm x)^{\frac{1-\lambda}{r}} \\
	 \overset{(a)}{\leq} \Big{(}\sum_{{\bm x}\in \mathcal{X}^{N}}f(\bm x)^{\frac{\lambda}{r}\frac{r}{\lambda s}}\Big{)}^{\frac{\lambda s}{r}} &  \Big{(}\sum_{{\bm x}\in \mathcal{X}^{N}}f(\bm x)^{\frac{1-\lambda}{r}\frac{r}{(1-\lambda)t}}\Big{)}^{\frac{(1-\lambda)t}{r}} \\
	  =  \Big{(}\sum_{{\bm x}\in \mathcal{X}^{N}}f(\bm x)^{\frac{1}{s}}\Big{)}^{\frac{\lambda s}{r}} & \Big{(}\sum_{{\bm x}\in \mathcal{X}^{N}}f(\bm x)^{\frac{1}{t}}\Big{)}^{\frac{(1-\lambda)t}{r}}
	\label{eq:D15_ch3}
\end{split}
\end{equation}
where the inequality (a) comes from the direct application of H\"{o}lder's inequality \cite{RyzhikBook}. Raising the left and right sides of Eq.~\eqref{eq:D15_ch3} to $r$ and taking the base-2 logarithm, we obtain the following result
\begin{equation}
\begin{split}
	r\log_{2}\sum_{{\bm x}\in \mathcal{X}^{N}}f(\bm x)^{\frac{1}{r}} &\leq  \lambda s  \log_{2}\Big{(}\sum_{{\bm x}\in \mathcal{X}^{N}}f(\bm x)^{\frac{1}{s}}\Big{)}  
	 + (1-\lambda) t \log_{2}\Big{(}\sum_{{\bm x}\in \mathcal{X}^{N}}f(\bm x)^{\frac{1}{t}}\Big{)}
\end{split}
	\label{eq:D16_ch3}
\end{equation}
which is equivalent to Eq.~\eqref{eq:D14_ch3} and proves that $F(r)$ is convex. The equality in Eq.~\eqref{eq:D15_ch3} holds, if and only if, $$\frac{f(\bm{x})^{\frac{1}{t}}}{f(\bm{x})^{\frac{1}{s}}} = f(\bm{x})^{\frac{1}{t}-\frac{1}{s}}= c, \quad  \forall {\bm{x}}\in {\mathcal{X}}^{N}$$ where $c$ is a constant, which is only feasible if $f({\bm{x}})$ is a uniform distribution. Thus, the function $F(r)$ is strictly convex unless $f({\bm{x}})$ is a uniform distribution.

By simple argumentation, it is now trivial to show that the function $E(\rho)$ is (strictly) convex. The first summand in Eq.~\eqref{eq:D12_ch3} is an affine function of $\rho$ and is convex. The second summand in Eq.~\eqref{eq:D12_ch3} is a composition of the (strictly) convex function $F(r)$ with the affine expression $1+\rho$, which yields a (strictly) convex function of $\rho$ \cite{convopt}. Hence, $E(\rho)$ is a convex function since it is a sum of convex functions, with strict convexity unless $f(\bm{x})$ is a uniform distribution.

\end{appendices}

\bibliographystyle{IEEEtran}
\bibliography{corsourcesbib}

\end{document}